\documentclass[a4paper,11pt,nofootinbib]{article}

\pdfoutput=1 % to allow compilation in JCAP
\usepackage{jcappub}
\usepackage{amsmath,amssymb}
\usepackage{float}
\usepackage{bm}
\usepackage{xcolor}
\usepackage{graphicx}
\usepackage{physics}
\usepackage{enumerate}
\usepackage{enumitem}
\usepackage{geometry}
\usepackage{xspace}
\usepackage{pdflscape}
\usepackage{placeins}
\usepackage{epstopdf}
\usepackage{comment}
\usepackage{subfig}

\makeatletter
\gdef\@fpheader{}
\g@addto@macro\bfseries{\boldmath}
\makeatother

\let\oldsqrt\sqrt
\def\sqrt{\mathpalette\DHLhksqrt}
\def\DHLhksqrt#1#2{%
\setbox0=\hbox{$#1\oldsqrt{#2\,}$}\dimen0=\ht0
\advance\dimen0-0.2\ht0
\setbox2=\hbox{\vrule height\ht0 depth -\dimen0}%
{\box0\lower0.4pt\box2}}

\newcommand{\sss}[1]{{\scriptscriptstyle{#1}}}
\newcommand{\boldmathsymbol}[1]{{\ensuremath{\boldsymbol{#1}}}}

\newcommand{\uPl}{\mathrm{Pl}}

\newcommand{\usssPl}{\sss{\uPl}}

\newcommand{\Mp}{M_\usssPl}

\newcommand{\beq}{\begin{equation}}
\newcommand{\eeq}{\end{equation}}
\newcommand{\bea}{\begin{equation}\begin{aligned}}
\newcommand{\eea}{\end{aligned}\end{equation}}

\newlength{\wsingfig}
\newlength{\wdblefig}
\newlength{\wquadfig}
\newlength{\wtriplefig}
\newcommand{\Eq}[1]{Eq.~(\ref{#1})}

\newcommand{\Fig}[1]{Fig.~{\ref{#1}}}

\newcommand{\Sec}[1]{Sec.~\ref{#1}}

\newcommand{\Hc}[1]{\mathcal{H}}
\renewcommand{\Hc}{\mathcal{H}}

\def\doi{http://doi.org}

\date{today}

\title{Gravitational waves from primordial black hole isocurvature: the effect of non-Gaussianities }
\author[a,b,c]{Xin-Chen He,}

\author[a,b,c]{Yi-Fu Cai,}

\author[a,b,c]{Xiao-Han Ma,}

\author[d,e,f]{Theodoros Papanikolaou,}

\author[f,a,b,g]{Emmanuel N. Saridakis,}

\author[h,i,j]{Misao Sasaki}

\affiliation[a]{Department of Astronomy, School of Physical Sciences, University of Science and Technology of China, Hefei, Anhui 230026, China}

\affiliation[b]{CAS Key Laboratory for Research in Galaxies and Cosmology, School of Astronomy and Space Science, University of Science and Technology of China, Hefei, Anhui 230026, China}

\affiliation[c]{Deep Space Exploration Laboratory, Hefei 230088, China}

\affiliation[d]{Scuola Superiore Meridionale, Largo San Marcellino 10, 80138 Napoli, Italy}

\affiliation[e]{Istituto Nazionale di Fisica Nucleare (INFN), Sezione di Napoli, Via Cinthia 21, 80126 Napoli, Italy}

\affiliation[f]{National Observatory of Athens, Lofos Nymfon, 11852 Athens, 
Greece}

\affiliation[g]{Departamento de Matem\'{a}ticas, Universidad Cat\'{o}lica del Norte, Avda. Angamos 0610, Casilla 1280 Antofagasta, Chile}

\affiliation[h]{Kavli Institute for the Physics and Mathematics of the Universe (WPI), UTIAS The University of Tokyo, Kashiwa, Chiba 277-8583, Japan}

\affiliation[i]{Center for Gravitational Physics, Yukawa Institute for Theoretical Physics, Kyoto University, Kyoto 606-8502, Japan}

\affiliation[j]{Leung Center for Cosmology and Particle Astrophysics,
National Taiwan University, Taipei 10617}

\emailAdd{xinchenhe@mail.ustc.edu.cn}
\emailAdd{yifucai@ustc.edu.cn}
\emailAdd{mxh171554@mail.ustc.edu.cn}
\emailAdd{t.papanikolaou@ssmeridionale.it}
\emailAdd{msaridak@noa.gr}
\emailAdd{misao.sasaki@ipmu.jp}

\abstract{ 
Ultra-light primordial black holes (PBHs) with masses $M_{\rm PBH}<5\times 10^8\mathrm{g}$ can dominate transiently the energy budget of the Universe and reheat the Universe through their evaporation taking place before Big Bang Nucleosynthesis. The isocurvature energy density fluctuations associated to the inhomogeneous distribution of a population of such PBHs can induce an abundant production of GWs due to second-order gravitational effects. In this work, we discuss the effect of primordial non-Gaussianity on the clustering properties of PBHs and study the effect of a clustered PBH population on the spectral shape of the aforementioned induced GW signal. In particular, focusing on local-type non-Gaussianity we find a double-peaked GW signal with the amplitude of the low-frequency peak being proportional to the square of the non-Gaussian parameter $\tau_\mathrm{NL}$. Remarkably, depending on the PBH mass $M_{\rm PBH}$ and the initial abundance of PBHs at formation time, i.e. $\Omega_\mathrm{PBH,f}$, this double-peaked GW signal can lie well within the frequency bands of forthcoming GW detectors, namely LISA, ET, SKA and BBO, hence rendering this signal falsifiable by GW experiments and promoting it as a novel portal probing the primordial non-Gaussianity. 
}

\begin{document}
\begin{flushleft}
YITP-24-93
\end{flushleft}

\maketitle

\section{Introduction}

Inflation is the most prevailing paradigm of the very early Universe that is able to address issues such as the horizon and flatness problems, while can also explain the origin of structures at different cosmological scales~\cite{Guth:1980zm, Linde:1981mu, Starobinsky:1980te}. Therefore, studying primordial cosmological perturbations at various scales is an effective way to examine the inflationary scenario. In particular, observations of the Cosmic Microwave Background (CMB) and Large Scale Structures (LSS) have already provided us with insights into large-scale primordial perturbations, reflecting the physics in the early stages of inflation~\cite{Planck:2018vyg, Lesgourgues:2007aa}. However, perturbations that exited the Hubble horizon during the later stages of inflation re-entered the Hubble horizon soon after the end of inflation, hence undergoing a longer period of non-linear evolution, something which makes them more challenging to observe directly. 

Primordial black holes (PBHs)~\cite{1967SvA....10..602Z, Carr:1974nx,Carr:1975qj}, formed by enhanced cosmological perturbations in the early Universe\footnote{The enhanced primordial energy density perturbations necessary for PBH formation can be naturally realized within for example single-field~\cite{Yokoyama:1998pt,Garcia-Bellido:2016dkw,Martin:2019nuw,Vennin:2020kng,Ragavendra:2020sop,Pattison:2021oen} and multi-field~\cite{GarciaBellido:1996qt,Kawasaki:1997ju,Frampton:2010sw,Clesse:2015wea,Inomata:2017okj,Pi:2017gih,Zhou:2020kkf,Wang:2024vfv} inflationary setups as well as within non-perturbative processes~\cite{Cai:2018tuh, Chen:2019zza,Chen:2020uhe,Cai:2021zsp,Cai:2022erk}.}, can serve as a crucial probe for studying primordial fluctuations at such small scales. They can acquire a wide range of masses depending on the time of their formation, having the potential to be the seeds for the supermassive black holes observed in the early Universe~\cite{Meszaros:1975ef, Bernal:2017nec, Bean:2002kx, 1984MNRAS.206..315C, Cai:2023ptf}, acting as well as viable candidates for dark matter~\cite{Chapline:1975ojl}. Furthermore, they can potentially source some of gravitational wave events detected by the LIGO/VIRGO collaboration~\cite{Sasaki:2016jop, LIGOScientific:2018mvr, LIGOScientific:2020ufj,Wang:2022nml} explaining as well the baryon asymmetry in the Universe~\cite{Baumann:2007yr, Hook:2014mla, Hamada:2016jnq, Carr:2019hud}. 

Within the standard General Relativity, PBHs possessing masses less than $10^{15}\mathrm{g}$ have already evaporated due to Hawking radiation. The Hawking evaporation of these light PBHs has implications for Big Bang Nucleosynthesis (BBN) and the CMB, with their abundance being constrained by current observations (See \cite{Sasaki:2018dmp, Carr:2020gox, Carr:2020xqk, Escriva:2022duf, LISACosmologyWorkingGroup:2023njw, Choudhury:2024aji} for comprehensive reviews). However, ultra-light PBHs with $M_{\rm PBH} < 10^{9}\mathrm{g}$, evaporated before BBN, thus escaping these observational constraints~\footnote{Recent literature has updated the constraints on such ultra-light PBHs by taking into account the memory burden effect~\cite{Jiang:2024aju,Thoss:2024hsr,Alexandre:2024nuo} or through BBN~\cite{Boccia:2024nly}.}. Such ultra-light PBHs, being naturally produced in the early Universe~\cite{Martin:2020fgl, Banerjee:2022xft, Papanikolaou:2023crz} can induce an early matter-dominated (eMD) phase before BBN~\cite{GarciaBellido:1996qt,Hidalgo:2011fj, Suyama:2014vga, Zagorac:2019ekv} with fruitful phenomenology~\cite{Inomata:2019ivs, Papanikolaou:2020qtd, Datta:2020bht, Hooper:2019gtx, Papanikolaou:2023oxq, Barrow:1990he, Bhaumik:2022pil, Bhaumik:2022zdd, Gehrman:2022imk,Domenech:2023mqk,Basilakos:2023jvp}. 

During the formation of PBHs~\footnote{Gravitational waves from the PBH evaporation were also studied in recent literature~\cite{Anantua:2008am, Dong:2015yjs, Domenech:2021wkk, Ireland:2023avg, Ireland:2023zrd}.}, the generation of induced gravitational waves (GWs) is an inevitable phenomenon~\cite{Matarrese:1992rp, Matarrese:1993zf, Matarrese:1997ay, Mollerach:2003nq}~(See here~\cite{Domenech:2021ztg} for a review). These GWs play a crucial role in probing such eMD phases and more importantly constraining the properties of ultra-light PBHs and the underlying fundamental physical theory. Intriguingly, for ultra-light PBHs, curvature perturbations are not the sole source of induced GWs. In particular, a population of PBHs can be modeled as a gravitationally interacting gas at large scales, being endowed with its own density perturbations~\cite{Papanikolaou:2020qtd, Domenech:2020ssp}. These density perturbations are isocurvature in nature at PBH formation time during early radiation dominated (eRD) era, and convert to adiabatic ones sourcing the gravitational potential after the onset of PBH domination. As evaporation occurs, the matter fluid that dominates the Universe and its perturbations convert into radiation. This transition results in oscillating fluctuations throughout the Universe, leading to observable induced GWs~\cite{Domenech:2023jve}. To the best of our knowledge, up to now there has been considered a Poisson-distributed PBH gas which is quite reasonable since if the primordial perturbations follow a Gaussian distribution, then PBHs, formed from enhanced density fluctuations exceeding a threshold $\delta_{\rm cr}$, are distributed randomly and can be aptly described by a Poisson distribution~\cite{Papanikolaou:2020qtd}. 

The two-point correlation function of PBHs, however, may deviate from the Poisson case if there exists a non-trivial higher order correlation function of the primordial density perturbations, leading to PBH clustering~\cite{Byrnes:2012yx, Young:2013oia, Pattison:2017mbe, Franciolini:2018vbk, Suyama:2019cst, Ezquiaga:2019ftu, Figueroa:2020jkf, Kitajima:2021fpq, Auclair:2024jwj, Animali:2024jiz}. In particular, as suggested in~\cite{Suyama:2019cst} the local-type trispectrum (four-point correlation function) of the primordial curvature perturbations can enhance the two-point correlation function of PBHs at super-Hubble scales. Notably, this mechanism was applied in~\cite{Papanikolaou:2024kjb} to illustrate that the product of the nonlinearity parameter at third order and the primordial power spectrum of curvature perturbations can be tightly bounded by combining the BBN and CMB observations as well as GW astronomy. 

In this work, we extend the study of~\cite{Papanikolaou:2024kjb} to a broadly scale-varying trispectrum case deriving at the end the GWs induced by the clustering effect of the non-Gaussian PBH gas. We remarkably find that the energy spectrum of the induced GWs is significantly enhanced at lower frequencies due to the non-vanishing trispectrum, forming a bi-peak structure. 

The paper is organised as follows. In Sec.~\ref{sec:PBH_matter_PS_Gaussian}, we revisit the paradigm of a Poisson-distributed PBH gas which originates from Gaussian density perturbations. Then, in Sec.~\ref{sec:PBH_matter_PS_non_Gaussian}, we incorporate local-type primordial non-Gaussianities into our framework, yielding a pronounced spatial clustering of PBHs. Treating this spatial distribution of PBHs as an isocurvature initial condition, we track the evolution of the associated gravitational potential throughout the eRD-eMD-lRD background evolution in Sec.~\ref{sec:PBH_Phi}, and calculate the large secondary GW signal produced by PBH evaporation in a semi-analytical way in Sec.~\ref{Sec:IGW}. Finally, we conclude in Sec.~\ref{sec:conclusion}.

%%%% Section 2: The Gaussian PBH matter power spectrum %%%%

\section{The Gaussian primordial black hole matter power spectrum}
\label{sec:PBH_matter_PS_Gaussian}

In this section, we consider a population (``gas") of PBHs randomly distributed in space which form at small scales in a radiation-dominated, homogeneous Universe\footnote{At large scales, primordial curvature perturbations provide the main contribution to the matter inhomogeneities but they are negligible compared to the ones generated by PBHs, at least in the range of scales we are interested in.}. For simplicity, we assume that PBHs form from enhanced primordial curvature perturbations which are characterised by a sharply peaked power spectrum. Therefore, the PBH mass function, defined as the energy density contribution of PBHs per logarithmic mass, is almost monochromatic. The initial PBH abundance at formation time is then given by~\cite{Carr:2020gox}
\begin{equation}
    \Omega_\mathrm{PBH,f}\equiv\frac{\rho_{\rm PBH,f}}{3H_{\rm f}^2\Mp^2}\,,
\end{equation}
where the subscript ``f'' refers to the PBH formation time $t_{\rm f}$. If the initial PBH abundance is sizeable, PBHs can dominate later the energy budget of the Universe since their abundance increases with the scale factor as $\Omega_\mathrm{PBH} \sim \rho_\mathrm{PBH}/\rho_\mathrm{r}\propto a^{-3}/a^{-4}\propto a$, as one evolves a matter component within a radiation background.

One can define as well at this point the wavenumber $k_{\rm f}$ as $k_{\rm f}\equiv a_{\rm f}H_{\rm f}$, being the wavenumber crossing the Hubble horizon at the time of formation. For a sharply peaked primordial spectrum, $k_{\rm f}$ roughly equals to the wavenumber at the peak $k_*$ and has a one-to-one correspondence with the PBH mass, 
\begin{equation}
    M_{\rm PBH,f}=\frac{4\pi\gamma \Mp^2}{H_{\rm f}}\,,
\end{equation}   
where the correction factor $\gamma$ represents the fraction of the horizon mass collapsing to PBHs. In general $\gamma$ depends on the equation-of-state parameter at the time of PBH formation and for radiation it is $\gamma \simeq 0.36$~\cite{Musco:2008hv}. 
Note that the formation of PBHs is a rare event arising from enhanced curvature perturbations, hence the mean separation distance between two PBHs is much larger than $k_{\rm f}^{-1}$. 

At distances much larger than the mean separation scale between PBHs, the PBHs can be effectively treated as a pressureless fluid. This fluid possesses density perturbations that can be analyzed within the framework of cosmological perturbation theory. Since PBHs form from high peaks of the density field, the spatial correlation of PBHs is expected to depend on the correlation of the primordial perturbations. For Gaussian primordial curvature perturbations, their peaks are uncorrelated, implying that PBHs are randomly distributed in space~\cite{Ali-Haimoud:2018dau,Desjacques:2018wuu}. Consequently, their statistics follow the Poisson distribution, resulting in a two-point correlation function for the PBH density contrast which reads as
\begin{equation}
    \langle\delta_{\rm PBH}({\bm x}_1)\delta_{\rm PBH}({\bm x}_2)\rangle=\frac{4\pi}{3}\left(\frac{\bar r}{a}\right)^3\delta_{\rm D}(\bm x_1-\bm x_2)\,,
\end{equation}
where $\bm x$ denotes comoving coordinates, and $\bar r$ represents the mean separation distance between two PBHs, given by $\bar r = (\tfrac{3M_{\rm PBH}}{4\pi\rho_{\rm PBH}})^{1/3}$. At scales smaller than $\bar r$, the description of PBH fluid fails, which gives an ultra-violet (UV) cutoff reading as~\cite{Papanikolaou:2020qtd}
\begin{equation}
    k_{\rm UV}\equiv \frac{a}{\bar r}=\frac{a_{\rm f}}{{\bar r}_{\rm f}}=\gamma^{1/3}\Omega^{1/3}_\mathrm{PBH,f}k_{\rm f},
\end{equation}
where $\Omega_\mathrm{PBH,f}$ is the PBH abundance at formation.
This UV cutoff represents actually the scale below which, one enters the non-linear regime, where perturbation theory breaks down.

In Fourier space, the reduced power spectrum of PBH density contrast defined as ${\cal P}(k)\equiv k^3|\delta_k|^2/(2\pi^2)$ can be straightforwardly expressed as
\begin{equation}\label{eq:ppoi}
    {\cal P}_{\delta_\mathrm{PBH}, \mathrm{G}}(k)=\frac{2}{3\pi}\left(\frac{k}{k_{\rm UV}}\right)^3\,.
\end{equation}
where the subscript ``G'' denotes the Gaussian nature of the primordial curvature perturbations.

%%%% Section 3: The PBH matter power spectrum in the presence of non-Gaussianities %%%%
\section{The primordial black hole matter power spectrum in the presence of non-Gaussianities}\label{sec:PBH_matter_PS_non_Gaussian}

Let us study now on the effects of primordial non-Gaussianities on the PBH power spectrum $ {\cal P}_{\delta_\mathrm{PBH}}(k)$. The effects of non-Gaussianities on PBH distribution can be divided into two aspects.
First, in the presence of primordial non-Gaussianities, different scale primordial curvature perturbation modes are now correlated. In particular, the energy density peak of a collapsing overdensity region with a characeristic size $R\sim H^{-1}_{\rm f}$ will be modulated by the long-wavelength modes that are on super-Hubble scales at PBH formation time~\cite{Tada:2015noa, Young:2015kda}. Consequently, the PBHs formed out of these rare, horizon-size, high peaks of the density field will have spatial correlations, leading to clustering~\cite{Ali-Haimoud:2018dau, Byrnes:2012yx, Young:2013oia, Franciolini:2018vbk, Ferrante:2022mui, Gow:2022jfb}. It is worth noting that there is no conflict with causality, as the high peaks of the density field separated by a super-Hubble distance do not interact directly with each other. Instead, they are each locally connected to the long-wavelength modes at their respective locations. These long-wavelength modes, in turn, are correlated with each other due to causal processes established early in inflation, when they were still within the Hubble horizon. Second, the initial abundance of PBHs can be modified by primordial non-Gaussianities~\cite{Young:2015kda,Franciolini:2018vbk,Matsubara:2022nbr}, resulting in a Universe filled with denser (or sparser) PBH gas equipped with a enhanced (or decreased) Poisson noise. 
Consequently, taking these considerations into account, the PBH matter power spectrum can be expressed as 
\beq\label{eq:P_PBH_full0}
    {\cal P}_{\delta_\mathrm{PBH}}(k)={\cal P}_{\delta_\mathrm{PBH},\mathrm{NG}}(k)+{\cal P}_{\delta_\mathrm{PBH},\mathrm{Poisson}}(k)\,,
\eeq
where the first and second term on the right hand side are the contribution from the clustering effect and the enhanced (or decreased) Poisson noise, respectively. Note that the Poisson term ${\cal P}_{\delta_\mathrm{PBH},\mathrm{Poisson}}$ should be same as \Eq{eq:ppoi} up to an overall factor, as the modification in PBH abundance effectively changes the size of $k_{\rm UV}$ while keeping the scaling unchanged. In contrast, the clustering term can significantly alter the scale dependence of the PBH matter power spectrum, providing a distinctive signature of non-Gaussianities.

In order to quantify such effects we choose to primarily focus on local-type non-Gaussianities. As widely adopted in the literature, one needs to introduce the non-linearity parameters $f_{\rm NL}$, $g_{\rm NL}$, and $\tau_{\rm NL}$ in order to quantify the relative amplitude of non-Gaussian features within the perturbative regime. The definitions of these non-linearity parameters are derived from the relations between the 3-point and 4-point correlation functions and the 2-point function and read as follows:
\boldsymbol{}
\begin{subequations}
\begin{align}
    \expval{\mathcal{R}(\vb*{k_1})\mathcal{R}(\vb*{k_2})} &\equiv (2\pi)^3\delta^{(3)}(\vb*{k_1}+\vb*{k_2})P_{\mathcal{R}}(k)\\
    \expval{\mathcal{R}(\vb*{k_1})\mathcal{R}(\vb*{k_2})\mathcal{R}(\vb*{k_3})} &\equiv (2\pi)^3\delta^{(3)}(\vb*{k_1}+\vb*{k_2} + \vb*{k_3}) \\
    &\quad \times \frac{6}{5}f_{\rm NL}
    \left[ P_{\mathcal{R}}(k_1)P_{\mathcal{R}}(k_2) + 2~\text{perms} \right] \nonumber \\
    \expval{\mathcal{R}(\vb*{k_1})\mathcal{R}(\vb*{k_2})\mathcal{R}(\vb*{k_3})\mathcal{R}(\vb*{k_4})} &\equiv (2\pi)^3\delta^{(3)}(\vb*{k_1}+\vb*{k_2} + \vb*{k_3} + \vb*{k_4}) \\
    &\quad \times\bigg\{ \frac{54}{25}g_{\rm NL}\left[P_{\mathcal{R}}(k_1)P_{\mathcal{R}}(k_2)P_{\mathcal{R}}(k_3) + 3~\text{perms} \right] \nonumber\\
    &\hspace{3em} + \tau_{\rm NL}
     \left[P_{\mathcal{R}}(k_1)P_{\mathcal{R}}(k_2)P_{\mathcal{R}}(\abs{\vb*{k_1}+\vb*{k_3}}) + 11~\text{perms} \right] \bigg\}. \nonumber
     \label{eq:4p}
\end{align}
\end{subequations}
It is worth noting that this definition does not assume these non-linearity parameters to be constant. They can in principle vary depending on the scale.

Following the approach developed in~\cite{Suyama:2019cst} and generalising it to the case of scale-dependent non-linearity parameters, one can derive the PBH power spectrum in the presence of local-type non-Gaussianities. After a lengthy calculation [See Appendix \ref{app:PBH matter power spectrum} for more details] one can derive the PBH power spectrum  ${\cal P}_{\delta_\mathrm{PBH}}(k)$ as
\begin{eqnarray}\label{eq:ppbhng}
    {\cal P}_{\delta_\mathrm{PBH}}(k)&\simeq&\left(\frac{4\nu}{9\sigma_R}\right)^4{\cal P}_{\mathcal{R}}(k)\int \frac{\mathrm{d}^3p_1 \mathrm{d}^3p_2}{(2\pi)^6}\tau_\mathrm{NL}(p_1,p_2)W^2_\mathrm{local}(p_1)W^2_\mathrm{local}(p_2)P_{\mathcal{R}}(p_1)P_{\mathcal{R}}(p_2)\cr\cr
    &&\;+\frac{k^3}{2\pi^2}\cdot(k\textrm{-independent terms})\cr\cr
    &\equiv&\nu^4\bar{\tau}_{\rm NL}{\cal P}_{\cal R}(k)+\frac{k^3}{2\pi^2}\cdot(k\textrm{-independent terms})\cr\cr
    &\equiv& {\cal P}_{\delta_\mathrm{PBH},\mathrm{NG}}(k)+{\cal P}_{\delta_\mathrm{PBH},\mathrm{Poisson}}(k)
    \,,
\end{eqnarray}
where $W_{\rm local}(k)\equiv(kR)^2W_R(k)$ is a smoothing function with a smoothing scale $R$, which basically smooths the energy density field over scales smaller than the Hubble scale avoiding in this way the formation of PBHs on small scales, namely the cloud-in-cloud problem~\cite{Ando:2018qdb,Young:2019osy}, removing as well scales larger than the scale $R$ to maintain causality. For a sharply peaked primordial spectrum, as the one studied in our scenario, $R$ is of the same order of $k_*^{-1}$. %Here, we choose the window function $W_R(k)=\exp(-k^2R^2/2)$. 
The peak height $\nu$ is defined as $\nu\equiv\delta_{\rm cr}/\sigma_{R}$, where $\delta_{\rm cr}$ is the critical value of density contrast for PBH formation and $\sigma_R$ is the standard deviation of the density contrast on the comoving slice. The parameter $\bar{\tau}_{\rm NL}$ in \Eq{eq:ppbhng} is an effective $\tau_{\rm NL}$ parameter defined as
\begin{eqnarray}\label{eq:eff_tau_NL}
\bar{\tau}_{\rm NL} \equiv \left(\frac{4}{9\sigma_R}\right)^4 \int \frac{\mathrm{d}^3p_1 \mathrm{d}^3p_2}{(2\pi)^6}\tau_\mathrm{NL}(p_1,p_2,p_1,p_2) W^2_\mathrm{local}(p_1)W^2_\mathrm{local}(p_2)P_{\mathcal{R}}(p_1)P_{\mathcal{R}}(p_2)\,, \nonumber\\
\end{eqnarray}
which equals $\tau_{\rm NL}$ when $\tau_{\rm NL}$ is a constant, as is usually considered. As mentioned above, a constant $\tau_{\rm NL}$ may not be valid on all scales. In general, one expects some non-trivial or nonlinear process at the late stages of inflation, generating non-Gaussian primordial fluctuations at small scales~\cite{Riotto:2010nh, Byrnes:2010ft,Martin:2012pe,Namjoo:2012aa}. As a phenomenological extension, in this work, we will assume a scale-dependent local-type non-linearity parameter $\tau_\mathrm{NL}=\tau_\mathrm{NL}(k_1,k_2,k_3,k_4)$, parametrizing it as a first phenomenological approximation with a log-normal distribution peaked at the scale of PBH formation $k_*$, i.e.
\begin{eqnarray}\label{eq:tau_NL_log_normal}
\tau_\mathrm{NL} (k_1,k_2,k_3,k_4)  = \frac{\tau_\mathrm{NL}(k_*)}{6}\left[e^{-\frac{1}{2\sigma_\tau^2}\left(\ln^2\frac{k_1}{k_*}+\ln^2\frac{k_2}{k_*}\right)}+\;5\;\mathrm{perms}\right]\,,
\end{eqnarray}
which is motivated by the case of a scale-dependent $f_{\rm NL}$. 
Note that in the equilateral limit, $k_1=k_2=k_3 = k_4 =k_*$, one gets $\tau_\mathrm{NL}$ at $k_*$. 

It is important to mention here that \Eq{eq:ppbhng} is derived in the large-scale limit, i.e. when $kR\ll 1$, which is consistent with the PBH fluid description described above, where the scale of our interest $k<k_{\rm UV}\ll k_*$. 
As one may see from \Eq{eq:ppbhng}, the first term on the right-hand side is proportional to the spectrum of primordial curvature perturbation, implying that if $\tau_{\rm NL}$-type non-Gaussianity exists, it will give rise to clustering of PBHs on large scales. Additionally, $\tau_{\rm NL}$ provides the leading order term in the clustering term, as can be seen in Appendix \ref{app:PBH matter power spectrum}, thereby contributing dominantly to the clustering behavior. It is important to emphasize that the ${\cal P}_{\delta_\mathrm{PBH},\mathrm{Poisson}}$ corresponding to the ``$k\textrm{-independent terms}$'' in \Eq{eq:ppbhng} also contains components related to the non-Gaussian parameters $f_\mathrm{NL}$, $g_\mathrm{NL}$ [See Appendix \ref{app:PBH matter power spectrum}], which can be interpreted as the effect of non-Gaussianities on the initial PBH abundance $\Omega_\mathrm{PBH,f}$ as mentioned above.

Interestingly, considering PBHs as a dark matter component of the Universe, one can treat their spatial distribution behaving as dark matter isocurvature perturbations~\cite{Tada:2015noa,Young:2015kda}. 
The tight constraints on the fraction of dark matter isocurvature perturbations from CMB observations~\cite{Planck:2018jri} can be translated into extremely tight upper bound limits on $\tau_{\rm NL}$~\cite{Suyama:2019cst}. {However, we need to note that in our scenario, the ultra-light PBHs evaporate into radiation very early, allowing ample time for thermal equilibrium to be established between particle species before BBN. As a result, all the particle species fluctuate together, leading to a vanishing isocurvature perturbation consistent with tight constraints from CMB.}

%%%%%%%%%%%%%%%%%%%%%% Section 4: The primordial black hole gravitational potential %%%%%%%%%%%%%%%%%%%%%%%%%%%%%%%%%%%%%%%%
\section{The primordial black hole gravitational potential}\label{sec:PBH_Phi}
On scales of interest, specifically where $k<k_{\rm UV}$, adiabatic fluctuations from inflation are relatively negligible. Consequently, we assume that the universe remains uniform in total energy density during the formation of PBHs.  However, the discrete and compact nature of PBHs introduces spatial variations in the relative energy density of different components, namely the radiation and the PBH fluids. Consequently, isocurvature perturbations, where the relative number density of species spatially fluctuates, arises from the formation of PBHs~\cite{Inman:2019wvr}. One then can introduce the isocurvature perturbation $S$ for a matter-radiation fluid system defined with respect to the energy density perturbations of each fluid as
\beq\label{eq:S_definition}
S = \frac{\delta\rho_\mathrm{PBH}}{\rho_\mathrm{PBH}} - \frac{3}{4} \frac{\delta\rho_\mathrm{r}}{\rho_\mathrm{r}}\,.
\eeq
At PBH formation time, since the energy density contribution of PBHs is negligible, compared to that of radiation, i.e. $\rho_\mathrm{PBH,f}\ll \rho_\mathrm{r,f}$, and the total energy is conserved, i.e. $\delta\rho_\mathrm{PBH,f}+\delta\rho_\mathrm{r,f} = 0$, one can show from \Eq{eq:S_definition} that 
\beq
S \simeq  \frac{\delta\rho_\mathrm{PBH,f}}{\rho_\mathrm{PBH,f}} \simeq  \frac{\delta n_\mathrm{PBH,f}}{n_\mathrm{PBH,f}}\,,
\eeq
where $n_\mathrm{PBH}$ is the PBH comoving number density. Thus, the initial isocurvature perturbation $S$ can be identified with the initial density contrast of PBH fluid, $\delta_\mathrm{PBH,f}$. 
These initial PBH isocurvature perturbations, will convert into adiabatic curvature perturbations associated with a PBH gravitational potential $\Phi$ \`a la Bardeen after PBHs dominate over radiation. 
The dynamics of $\Phi$ is governed by the following equation:
\bea
\label{eq:Phi_evolution}
\Phi_\boldmathsymbol{k}^{\prime\prime} + 3\mathcal{H}(1+c_s^2)\Phi_\boldmathsymbol{k}^{\prime} +\left[\mathcal{H}^2(1+3c_s^2)+2\mathcal{H}'\right]\Phi_\boldmathsymbol{k}+c_s^2k^2\Phi_\boldmathsymbol{k} =\frac{a^2\rho_\mathrm{PBH}}{2\Mp^2}c_s^2S\,,
\eea
where $\mathcal{H}\equiv a^\prime/a$ is the comoving Hubble parameter, the prime here stands for differentiation with respect to the conformal time $\eta$ defined as $\mathrm{d}t \equiv a\mathrm{d}\eta$, and the sound speed $c_s^2$ is defined as 
\beq\label{eq:cs}
c_s^2 = \frac{4}{9}\frac{\rho_\mathrm{r}}{\rho_\mathrm{PBH}+\frac{4}{3}\rho_\mathrm{r}}\,,
\eeq
varying smoothly from $1/3$ in the eRD era to $0$ deep within the eMD era driven by PBHs. 

Since the value of $\Phi$ depends on when the mode enters the Hubble horizon, we analyze the modes that enter the Hubble horizon during eRD and eMD era separately to obtain the value of $\Phi$ deep in the eMD era. We denote the time when PBHs start to dominate the Universe as $t_{\rm d}$, the value of the scale factor and the wavenumber of the mode crossing the Hubble horizon at $t_{\rm d}$ as $a_{\rm d}$ and $k_{\rm d}=a_{\rm d}H_{\rm d}$, respectively.
For modes entering the Hubble horizon during eMD, namely $k\ll k_{\rm d}$, the super-Hubble evolution of $\Phi$ can be obtained by solving \Eq{eq:Phi_evolution} in the long-wavelength limit and choosing the isocurvature initial condition, namely $\Phi\to 0$ as $a\ll a_{\rm d}$. Consequently, the non-decaying mode of $\Phi$ during eMD when $a\gg a_{\rm d}$ is 
\bea\label{Phi_super_horizon}
\Phi_{\bm k}\simeq \frac15 \delta_\mathrm{PBH,f} \quad \mathrm{for}\quad k\ll {k_{\rm d}}\,,
\eea
which remains constant after the mode enters the Hubble horizon. 

On the other hand, for modes entering the Hubble horizon during the eRD era, namely $k\gg k_{\rm d}$, the gravitational potential $\Phi$ is first suppressed in proportion to $a^{-2}$ until the eMD era, then remains constant. One can, in principle, solve \Eq{eq:Phi_evolution} using WKB approximation. However, since the energy density of radiation is negligible deep in the eMD era, instead of solving \Eq{eq:Phi_evolution}, it is more convenient to relate the sub-Hubble modes of $\Phi$ to the PBH density contrast through the Poisson equation, i.e.
\begin{equation}\label{eq:poieq}
    \Phi_{\bm k}=\frac{3}{2}\left(\frac{\cal H}{k}\right)^2\delta_{\rm PBH}\,,
\end{equation}
where the growth of the PBH density perturbations can be determined by the M\'esz\'aros growth equation~\cite{Meszaros:1974tb}
\beq\label{eq:Meszaros}
    \frac{{\rm d}^2 \delta_{\rm PBH}}{{\rm d} s^2}+\frac{2+3s}{2s(s+1)}\frac{{\rm d} \delta_{\rm PBH}}{{\rm d} s}-\frac{3}{2s (s+1)}\delta_{\rm PBH}=0\,, \quad \text{with}\quad s\equiv\frac{a}{a_{\rm d}}\,.
\eeq
The non-decaying solution of \Eq{eq:Meszaros} gives rise to $\delta_{\rm PBH}\simeq 3s \delta_{\rm PBH}(t_{\rm f})/2$. 
Plugging it then in \Eq{eq:poieq}, we obtain $\Phi$ inside the Hubble horizon deep in the eMD era
\begin{equation}\label{Phi_sub_horizon}
    \Phi_{\bm k}\simeq\frac{9}{8}\left(\frac{k_{\rm d}}{k}\right)^2\delta_{{\rm PBH},{\rm f}}\quad \text{for}\quad k\gg k_{\rm d}\,.
\end{equation}
To conclude, using a crude interpolation between \Eq{Phi_super_horizon} and \Eq{Phi_sub_horizon}, we obtain the gravitational potential deep inside the eMD era when $a\gg a_{\rm d}$, 
\begin{equation}\label{eq:Phi_eMD}
    \Phi_{\rm eMD}(k)\equiv T_{\Phi, \rm eMD}(k)S=\left(5+\frac{8}{9}\frac{k^2}{{k}^2_\mathrm{d}}\right)^{-1} \delta_{{\rm PBH},{\rm f}}\,,
\end{equation}
where $S=\delta_{\rm PBH,f}$ and the transfer function $T_{\Phi,\rm eMD}(k)$ evolves the initial isocurvature perturbation $S$ to the adiabatic perturbation $\Phi$ during the eMD era. 

Since we focus on the GW signal induced by large oscillating acoustic waves of radiation, we need to determine the gravitational potential $\Phi$ right after PBH evaporation. When evolving $\Phi$ from the eMD era to the lRD era, one can solve \Eq{eq:Phi_evolution} under the instantaneous evaporation approximation~\cite{Domenech:2020ssp}. For the scales of our interest, i.e. scales deep inside the Hubble horizon at evaporation $x_{\rm evap}\equiv k\eta_{\rm evap}\gg 1$, $\Phi_{\rm lRD}^{\rm instant}$ can be approximated as 
\begin{equation}\label{eq:Phi_ins}
    \Phi_{\rm lRD}^{\rm instant}(k, \eta)= \left(\frac{x_{\rm evap}}{2\bar{x}}\right)^2\cos\left[c_s\left(\bar{x}-{x_{\rm evap}}/{2}\right)\right]\cdot\Phi_{\rm eMD}(k)~,
\end{equation}
where $\bar{x}\equiv k(\eta-\eta_{\rm evap}/2)$. 
One should also take into account that scales which have a time variation larger than the PBH evaporation rate $\Gamma\equiv - {\rm d}\ln M_{\rm PBH}/{\rm d}t$, i.e. $k/a\gg \Gamma$ are effectively suppressed by the non-zero pressure of the radiation fluid~\cite{Inomata:2020lmk}. This gives rise to an extra suppression factor $S_\Phi(k)$ in front of the gravitational potential $\Phi$ which reads as\footnote{More precisely, $\Phi$ at a given scale $k$ decouples from the fluctuations in the PBH energy density at some $t_{\rm dec}(k)\lesssim t_{\rm evap}$ when $a\Gamma|_{t=t_{\rm dec}(k)}\simeq k$~\cite{Inomata:2020lmk}. Before this, perturbations in the radiation fluid are not effectively produced and $\Phi$ remains sourced by the PBH fluctuation, i.e $k^2\Phi\sim a^2\rho_{\rm PBH}\delta_{\rm PBH}$. At the end, using the energy continuity equation of the PBH fluid, one can obtain that for $k/a\gg\Gamma$, $\Phi$ approximately decays as $\rho_{\rm PBH}\propto (1-t/t_{\rm evap})^{1/3}$. Thus, $\Phi$ gets suppressed by a factor of approximately $(1-t_{\rm dec}(k)/t_{\rm evap})^{1/3}\simeq (k/k_{\rm evap})^{-1/3}$ before decoupling from the PBH fluctuations. }
\beq\label{eq:supfactor}
S_\Phi(k)\equiv\frac{\Phi_{\rm lRD}}{\Phi_{\rm lRD}^{\rm instant}}\simeq\left(\sqrt{\frac{2}{3}}\frac{k}{k_\mathrm{evap}}\right)^{-1/3}\,\,,
\eeq
where $k_{\rm evap}$ is the wavenumber of the mode crossing the Hubble horizon at the time of evaporation $\eta_{\rm evap}$. This $k^{-1/3}$ suppression factor does not apply for modes with wavelengths $k<k_{\rm evap}$ which are superhorizon at the time of evaporation. These modes are unaffected by PBH evaporation remaining constant, and decay as $\eta^{-2}$ after reentering the horizon~\cite{Mukhanov:1990me}.

Finally, we obtain the gravitational potential during the lRD era, 
\begin{equation}\label{eq:Phi_lRD}
    \Phi_{\rm lRD}(k,\eta)=T_{\Phi}(k,\eta) S\,,
\end{equation}
where the transfer function $T_{\Phi}(k,\eta)$ includes the conversion from the isocurvature fluctuation to the adiabatic perturbation $\Phi$ as \Eq{eq:Phi_eMD}, and further evolution of $\Phi$ to the lRD era when induced GWs are abundantly produced: 
\begin{equation}\label{eq:T_Phi}
    T_{\Phi}(k,\eta)=S_{\Phi}(k)\left(\frac{x_{\rm evap}}{2\bar{x}}\right)^2\cos\left[c_s\left(\bar{x}-{x_{\rm evap}}/{2}\right)\right]\,T_{\Phi, \rm eMD}(k)
    \,.
\end{equation}
Thus, the reduced power spectrum for $\Phi$ during the lRD era is given by
\beq\label{eq:P_Phi_lRD}
\mathcal{P}_{\Phi,\rm lRD}(k,\eta)=T_{\Phi}^2(k,\eta)\mathcal{P}_S(k)\,,
\eeq
where $\mathcal{P}_{S}(k)=\mathcal{P}_{\delta_\mathrm{PBH}}(k)$ is given by \Eq{eq:ppbhng}. 

\begin{figure}[t!]
\begin{center}
\includegraphics[width=0.9\textwidth]{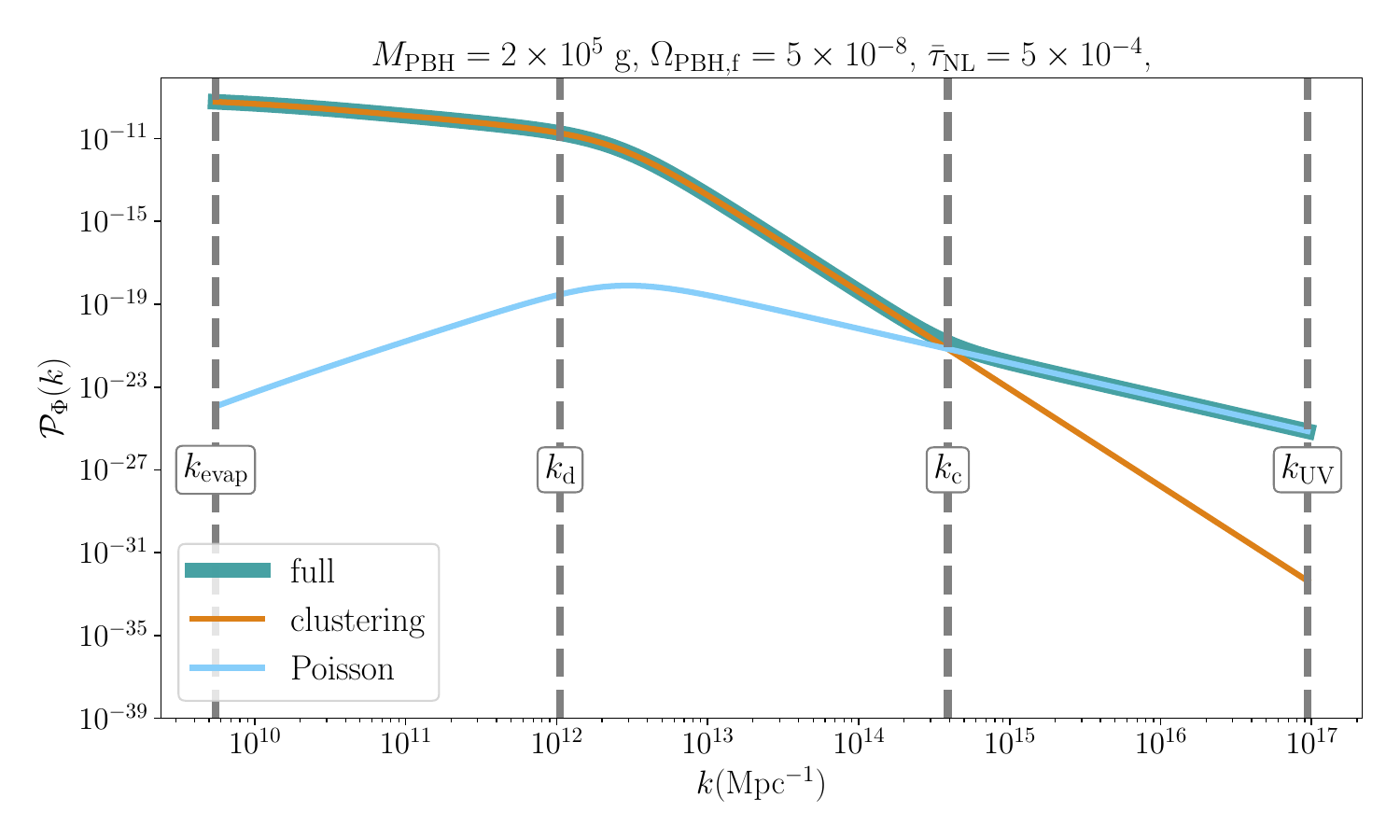}
\includegraphics[width=0.9\textwidth]{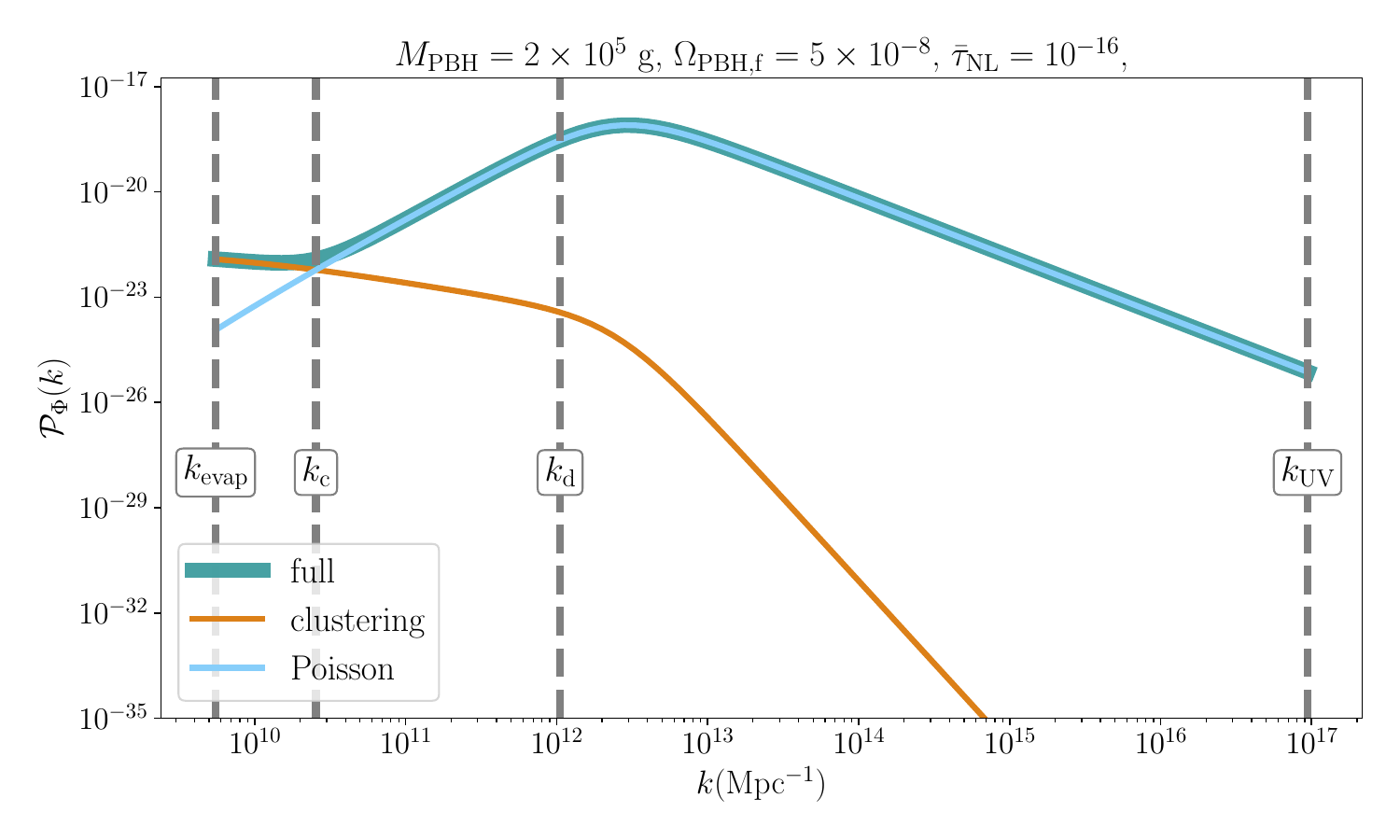}
\caption{{The power spectrum of the PBH gravitational potential for $M_\mathrm{PBH}= 5\times 10^{5}\mathrm{g}$ and $\Omega_\mathrm{PBH,f}=5\times 10^{-8}$ for $\tau_\mathrm{NL}=5\times 10^{-4}$ (Upper Panel) and for $\tau_\mathrm{NL}=10^{-16}$ (Lower Panel). The solid orange curves stand for the Poisson contribution to $\mathcal{P}_{\Phi}(k)$ while the solid blue ones denote the non-Gaussian clustering modification of $\mathcal{P}_{\Phi}(k)$. The green thick line refers to the full $\mathcal{P}_{\Phi}(k)$.}}
\label{fig:P_PBH_vs_tau_NL}
\end{center}
\end{figure}

To illustrate the spectral shape of the gravitational potential by varying the trispectrum parameter $\tau_\mathrm{NL}$, we depict in \Fig{fig:P_PBH_vs_tau_NL} the power spectrum of gravitational potential right after PBH evaporation, denoted as  
\begin{eqnarray}\label{eq:P_Phi}
    \mathcal{P}_{\Phi}(k)\equiv {\cal P}_{\Phi,\rm lRD}(k,\eta_{\rm evap})=S^2_\Phi(k)\left(5+\frac{8}{9}\frac{k^2}{{k}^2_\mathrm{d}}\right)^{-2}\mathcal{P}_{\delta_\mathrm{PBH}}(k)\,.
\end{eqnarray}
Having fixed the value of the PBH mass to $M_\mathrm{PBH}= 5\times 10^{5}\mathrm{g}$ and $\Omega_\mathrm{PBH,f} = 5\times 10^{-8}$ respectively we show in \Fig{fig:P_PBH_vs_tau_NL} the gravitational potential ${\cal P}_\Phi(k)$ for two characteristic values of the non-linearity parameter $\tau_\mathrm{NL}$, namely $\tau_\mathrm{NL} = 10^{-18}$ (Left Panel) and $\tau_\mathrm{NL} = 10^{-4}$ (Right Panel). 

As one may see from \Fig{fig:P_PBH_vs_tau_NL}, we have the appearance of four characteristic scales of our problem at hand. These scales correspond to the modes crossing the Hubble horizon at the onset of the PBH-domination time and at the PBH evaporation time, i.e. $k_\mathrm{d}$ and $k_\mathrm{evap}$ respectively, the UV cut-off scale $k_\mathrm{UV}$ introduced in \Sec{sec:PBH_matter_PS_Gaussian} and the critical scale $k_\mathrm{c}$ below which the PBH power spectrum is dominated by the Poissonian term ${\cal P}_{\delta_\mathrm{PBH},\mathrm{Poisson}}$ in \Eq{eq:ppbhng}. $k_\mathrm{d}$, $k_\mathrm{evap}$ and $k_\mathrm{UV}$ are related explicitly to $k_\mathrm{f}$ and $\Omega_\mathrm{PBH,f}$ as follows~\cite{Domenech:2020ssp}:
\beq\label{eq:k_evap_d_UV}
\frac{k_\mathrm{evap}}{k_\mathrm{f}} = \left(\frac{3.8g_{*}\Omega_\mathrm{PBH,f}}{960\gamma}\right)^{1/3}\left(\frac{M_\mathrm{PBH,f}}{\Mp}\right)^{-2/3},\quad \frac{k_\mathrm{UV}}{k_\mathrm{f}} = \left(\frac{\Omega_\mathrm{PBH,f}}{\gamma}\right)^{1/3}, \quad \frac{k_\mathrm{d}}{k_\mathrm{f}} = \sqrt{2} \Omega_\mathrm{PBH,f}.
\eeq
Taking into account the evolution of the background Universe, the relation between $k_{\rm f}$ and initial parameters of PBHs can be expressed as 
\beq
\frac{k_\mathrm{f}}{10^{20} \mathrm{Mpc}^{-1}} \simeq \left(\frac{3.8\,g_{*}}{960\,\gamma^{3/4}}\right)^{2} \left(\frac{M_\mathrm{PBH,f}}{10^4\mathrm{g}}\right)^{-5/6}\left(\frac{\Omega_\mathrm{PBH,f}}{10^{-7}}\right)^{-1/3},
\eeq
where $g_{*}$ is the effective number of relativistic species at the epoch of PBH formation, being of the order of $100$ for PBHs evaporating before BBN~\cite{Carr:2020gox}. It is worth noting that the scale $k_{\rm f}=a_{\rm f} H_{\rm f}$ is a function of not only the PBH mass but also the PBH abundance, as both of them are needed in determining the onset and the duration of the eMD era.

From \Eq{eq:k_evap_d_UV} one can derive the hierarchy of the different scales at hand which will read as 
\beq
k_\mathrm{evap}<k_\mathrm{d}<k_\mathrm{UV}<k_\mathrm{f}.
\eeq
One then discriminates between two characteristic regimes depending on the value of the critical scale $k_\mathrm{c}$. 
For $k_\mathrm{d}<k_\mathrm{c}<k_\mathrm{UV}$ (Left Panel of \Fig{fig:P_PBH_vs_tau_NL}), one gets the following scaling for $\mathcal{P}_{\Phi}(k)$:
\begin{itemize}
\item{
$k_\mathrm{evap}<k_\mathrm{d}<k_\mathrm{c}<k_\mathrm{UV}<k_\mathrm{f}$
\beq \label{eq:scale2}
\mathcal{P}_{\Phi}(k) \propto  
\begin{cases}
   k^{-2/3},\quad k_\mathrm{evap}<k<k_\mathrm{d} \\
   k^{-14/3},\quad k_\mathrm{d}<k<k_\mathrm{c} \\
   k^{-5/3} ,\quad k_\mathrm{c}<k<k_\mathrm{UV}
\end{cases}
\eeq
}
\end{itemize}
However, for extremely small values of $\tau_\mathrm{NL}$ at $k_\mathrm{f}$ one is met with regime where $k_\mathrm{evap}<k_\mathrm{c}<k_\mathrm{d}$ (Right Panel of \Fig{fig:P_PBH_vs_tau_NL}). In this regime, we obtain the following scaling behaviour for $\mathcal{P}_{\Phi}(k)$:
\begin{itemize}
\item{$k_\mathrm{evap}<k_\mathrm{c}<k_\mathrm{d}<k_\mathrm{UV}<k_\mathrm{f}$
\beq \label{eq:scale1}
\mathcal{P}_{\Phi}(k) \propto  
\begin{cases}
   k^{-2/3},\quad k_\mathrm{evap}<k<k_\mathrm{c} \\
   k^{7/3},\quad k_\mathrm{c}<k<k_\mathrm{d} \\
   k^{-5/3} ,\quad k_\mathrm{d}<k<k_\mathrm{UV}
\end{cases}
\eeq
}
\end{itemize}

%%%%%%%%%%%%%%%%%%%% Section 5: Induced GWs %%%%%%%%%%%%%%%%%%%%%%%%%%%%%%%%%%%%%%%%%%%%%%%%%%%%%%%%
\section{Induced gravitational waves from primordial black hole isocurvature perturbations}\label{Sec:IGW}
In our previous discussion, we investigated isocurvature perturbations arising from the PBH number-density fluctuations due to the formation of PBHs in presence of non-Gaussianities, as well as how these initial isocurvature fluctuations convert into adiabatic perturbations, $\Phi$. Having extracted before the power spectrum of the PBH gravitational potential in presence of non-Gaussianities, let us study now the gravitational waves induced by scalar perturbations due to second order gravitational interactions [See~\cite{Domenech:2021ztg,Yuan:2021qgz} for a review]. 

\subsection{The basics of scalar induced gravitational waves}

Working in the Newtonian gauge, the perturbed metric can be written as~\footnote{The tensor perturbations at linear order are not considered here.}   
\bea
\label{metric decomposition with tensor perturbations}
\mathrm{d}s^2 = a^2(\eta)\left\lbrace-(1-2\Phi)\mathrm{d}\eta^2  + \left[(1+2\Phi)\delta_{ij} + h_{ij}\right]\mathrm{d}x^i\mathrm{d}x^j\right\rbrace.
\eea
Then, by Fourier transforming the tensor perturbations and minimising the third order gravitational action, one obtains the equation of motion for the tensor modes $h_\boldmathsymbol{k}$ which reads as
\beq
\label{Tensor Eq. of Motion}
h_\boldmathsymbol{k}^{s,\prime\prime} + 2\mathcal{H}h_\boldmathsymbol{k}^{s,\prime} + k^{2} h^s_\boldmathsymbol{k} =  {\cal S}^s_\boldmathsymbol{k}\, ,
\eeq
where $s = (+), (\times)$ denote the two polarisation states of the tensor modes in GR and $^\prime$ stands for differentiation with respect to the conformal time $\eta$. After some algebra, one can show that in a matter-radiation  Universe, the source function ${\cal S}^s_\boldmathsymbol{k}$ is given by~\cite{Ananda:2006af,Baumann:2007zm,Kohri:2018awv,Espinosa:2018eve}
\begin{eqnarray}
    \label{eq:Source:def}
{\cal S}^{s}_\boldmathsymbol{k}  &=& 4\int\frac{\mathrm{d}^3 \boldmathsymbol{q}}{(2\pi)^{3}}e^s_{ij}(\boldmathsymbol{k})q_i q_j\left[\Phi_\boldmathsymbol{q}\Phi_\boldmathsymbol{k-q} + \frac{3\rho}{2\rho_r}c_s^2\left(\frac{\Phi_\boldmathsymbol{q} ^{\prime}}{\mathcal{H}}+\Phi_\boldmathsymbol{q}\right)\left(\frac{\Phi_\boldmathsymbol{k-q} ^{\prime}}{\mathcal{H}}+\Phi_\boldmathsymbol{k-q}\right) \right]\cr\cr
&=& 4\int\frac{\mathrm{d}^3 \boldmathsymbol{q}}{(2\pi)^{3}}e^s_{ij}(\boldmathsymbol{k})q_i q_j\,{S_{\bm q}} {S_{\bm {k-q}}}F(\eta,q,|\bm{k-q}|)
\end{eqnarray}
where $c_s^2$ is given by \Eq{eq:cs} and $e^{s}_{ij}(k)$ are the polarization tensors~\cite{Baumann:2007zm}. Here, the second equality is obtained by substituting the transfer function $\Phi_{\bm k}=T_\Phi(k,\eta)S_{\bm k}$ given by \Eq{eq:Phi_lRD} and the function $F(\eta,q,|\bm{k-q}|)$ that contains information of evolution of $\Phi$ is expressed as
\begin{eqnarray}\label{eq:ffunc}
&&F(\eta,q,|\bm{k-q}|)=
T_\Phi(q\eta)T_\Phi(|\bm{k-q}|\eta)\cr\cr
&&\qquad\qquad\quad+\frac{3\rho}{2\rho_r}c_s^2\left(\frac{T_\Phi'(q\eta)}{{\cal H}}+T_\Phi(q\eta)\right)\left(\frac{T_\Phi'(|\bm{k-q}|\eta)}{{\cal H}}+T_\Phi(|\bm{k-q}|\eta)\right)\,.
\end{eqnarray}
In our setup, the initial condition at PBH formation time is isocurvature on our interested scales, as discussed above, hence it is the isocurvature perturbation $S$ that appears in  \Eq{eq:Source:def} instead of the primordial curvature perturbation ${\cal R}$ as usually considered.  

Solving \Eq{Tensor Eq. of Motion} with the use of Green's function formalism, one can show after a long but straightforward calculation that the power spectrum of the second order tensor modes can be expressed as 
\begin{equation}\label{eq:PhPS}   \overline{\mathcal{P}_h(k,\eta)}=8\int_0^{\infty}dv\int_{|1-v|}^{1+v}du\left[\frac{4v^2-(1+v^2-u^2)^2}{4uv}\right]^2\mathcal{P}_S(kv)\mathcal{P}_S(ku)\overline{I^2}(x,u,v)~;
\end{equation}
\begin{equation}\label{eq:Idef}  I(x,u,v)=\int_{x_i}^{x}d\tilde{x}\,G(x,\tilde{x})f(\tilde{x},u,v)~, 
\end{equation}
where ${\cal P}_S(k)$ is the initial PBH isocurvature perturbation power spectrum, equal to ${\cal P}_{\delta_{\rm PBH}}$ in \Eq{eq:ppbhng}, and the overbar in \Eq{eq:PhPS} represents the time average over several periods of oscillation. The function $f(x,u,v)$ in \Eq{eq:Idef} is obtained by replacing the arguments of \Eq{eq:ffunc} with $v\equiv q/k$, $u\equiv |\bm{k-p}|/k$, and $x\equiv k\eta$. The lower limit $x_i$ of integral in \Eq{eq:Idef} is related to the initial time and we choose it to be $x_i=x_{\rm f}\equiv k\eta_{\rm f}$ for consistency. 
At this point, one should emphasize that the kernel function $I(x,u,v)$ can be split in three pieces as
\beq
I(x,u,v) = I_\mathrm{eRD}(x,u,v,x_\mathrm{f},x_\mathrm{d}) +  I_\mathrm{eMD}(x,u,v,x_\mathrm{d},x_\mathrm{evap}) +  I_\mathrm{lRD}(x,u,v,x_\mathrm{evap})\,,
\eeq
where we write explicitly the dependence on $x_\mathrm{f}$, $x_\mathrm{d}$ and $x_\mathrm{evap}$ by requiring continuity of the tensor perturbations and their derivatives at $\eta_\mathrm{d}$ and $\eta_\mathrm{evap}$. Out of the above three terms, the last one, related to the lRD era, is the dominant one due to the enhanced amplitude of the oscillations of $\Phi$ during the lRD era~\cite{Inomata:2019ivs}. Furthermore, due to the sudden transition from the eMD era to the lRD era, valid for monochromatic PBH mass distributions, as the ones considered here, one expects no correlation between perturbation modes generated in different eras~\cite{Inomata:2019ivs}. Thus, oscillation average cross-product terms such as $\overline{I_\mathrm{eMD}I_\mathrm{lRD}}$, can be safely neglected~\cite{Domenech:2020ssp}. At the end, the average kernel squared can be approximately replaced by $\overline{I_{\rm lRD}^2}$ and the oscillation average tensor power spectrum can be recast  as
\beq\label{eq:phplrd}
\overline{\mathcal{P}_h(k,\eta)}\simeq\overline{\mathcal{P}_{h,\mathrm{lRD}}(k,\eta)}\,.
\eeq

\subsection{The kernel function $I(u,v,x)$}
When computing now the kernel $I_\mathrm{lRD}$, one should take into account the fact that in the case of a sudden transition from the eMD to the lRD era the dominant term in \Eq{eq:ffunc} is the one containing derivatives of the gravitational potential $\Phi$, namely the $\mathcal{H}^{-2}\Phi^{\prime 2}$ term. This is because the derivative term $\mathcal{H}^{-1}\Phi^\prime$ goes quickly during the transition from zero, $\mathcal{H}^{-1}\Phi^\prime_\mathrm{eMD}(\eta_{\rm evap})=0$ to $\mathcal{H}^{-1}\Phi^\prime_\mathrm{lRD}(\eta_{\rm evap})\sim k\eta_{\rm evap}\Phi_\mathrm{eMD}\gg \Phi_\mathrm{eMD}$ since  $k\eta_{\rm evap}\gg 1$ - we consider modes being deep in the sub-Hubble regime at PBH evaporation time. At the end, one can recast the kernel function $I(u,v,x)$ in the lRD era as 
\begin{equation}\label{eq:IlRD}
    I_{\rm lRD}(x,u,v,x_{\rm evap})\simeq\frac{1}{2}uv\int_{x_{\rm evap}/2}^{\bar{x}}\mathrm{d}\tilde{\bar{x}}\,\tilde{\bar{x}}^2\,G(\bar{x},\tilde{\bar{x}})\frac{\mathrm{d} T_{\Phi}(u\tilde{\bar{x}})}{\mathrm{d}(u\tilde{\bar{x}})}\frac{\mathrm{d} T_{\Phi}(v\tilde{\bar{x}})}{\mathrm{d}(v\tilde{\bar{x}})} \,,
\end{equation}
where $G(\bar x,{\tilde {\bar x}})=\frac{{\tilde {\bar x}}}{\bar x}\sin(\bar x- {\tilde {\bar x}})$ is the Green's function during the lRD era. From \Eq{eq:T_Phi}, one can show that for the scales of interest, i.e. $x_{\rm evap}\gg1$, the time derivative of the transfer function is given by
\begin{align}\label{eq:tprime}
\frac{\mathrm{d}}{\mathrm{d}(v \tilde{\bar x})}T_{\Phi}(v\tilde{\bar x})\simeq -c_s\left(\frac{x_{\rm evap}}{2\tilde{\bar x}}\right)^2\sin\left[c_s v\left(\tilde{\bar x}-x_{\rm evap}/2\right)\right]\,T_{\Phi,\rm eMD}(v\tilde{\bar x})\,,
\end{align}
where we have considered the limit of large $x_\mathrm{evap}$.
Working now in the late time limit $\bar x \gg 1$, we can safely set the upper limit of the integral in \Eq{eq:IlRD} to $\bar{x}\to\infty$. At the end, \Eq{eq:IlRD} can be recast as
\begin{equation}\label{eq:kernelRD2} 
    I_{\rm lRD}(\bar{x},u,v,x_{\rm evap})\simeq\frac{c_s^2 uv}{32\bar x}x_{\rm evap}^4{\cal I}(\bar x,u,v,x_{\rm evap})\,,
\end{equation}
where
\begin{equation}\label{eq:sint}
    {\cal I}(\bar x,u,v,x_{\rm evap})=\int_{0}^{\infty} \frac{\mathrm{d}x_2}{x_2+x_{\rm evap}/2}\sin(x_1-x_2)\sin(vc_s x_2)\sin(uc_s x_2).
\end{equation}
For simplicity we have defined $x_1\equiv\bar x-x_{\rm evap}/2$ and $x_2\equiv\tilde{\bar x}-x_{\rm evap}/2$.

The integral in \Eq{eq:sint} can be approximated analytically in two regimes, which give the dominant contributions to the GW signal~\cite{Inomata:2019ivs,Inomata:2020lmk}. In particular, one can show that \Eq{eq:sint} diverges when the frequency of the faster oscillating mode equals to the sum of frequency of the slower oscillating modes, that is when $u+v=c_s^{-1}$. Such resonance is typical of systems where two slow modes source a third one which propagates faster~\cite{Domenech:2021ztg}. Focusing therefore on these resonant modes, one finds that the resonant part of the average kernel function squared is approximately given by~\cite{Inomata:2019ivs,Domenech:2020ssp}
\begin{align}\label{eq:irdapp}
\overline{I^2_{\rm lRD,{\rm res}}}&(u+v\sim c_s^{-1},x_{\rm evap})\simeq \frac{c_s^4u^2v^2}{2^{15}\bar x^2}x_{\rm evap}^8{\rm Ci}^2(|1-(u+v)c_s|x_{\rm evap}/2)\,.
\end{align}

The second relevant contribution to the kernel comes from the peak in the integrand at $u\sim v\gg1$. Using that ${\rm Ci}(x\to\infty)\to0$ and ${\rm  Si}(x\to\infty)\to\pi/2$ we select the cosine and sine integral without $v$ dependences. In this case, the large $v$ contribution to the averaged kernel squared reads as
\begin{align}\label{eq:irdapp2}
\overline{I^2_{\rm lRD,{\rm LV}}}(u\sim v\gg 1,x_{\rm evap})\simeq&\frac{c_s^4u^2v^2}{2^{13}\bar{x}^2}x_{\rm evap}^8\Big[{\rm Ci}^2(x_{\rm evap}/2)+\big({\rm Si}(x_{\rm evap}/2)-\pi/2\big)^2\Big]\,.
\end{align}
Note that both kernels \eqref{eq:irdapp} and \eqref{eq:irdapp2} peak at $k=k_{\rm UV}$ where $x_{\rm evap}$ attains its maximum value. %Also notice that since we approximated the transfer function at the end of the eMD to be a broken power-law \eqref{eq:conversionfinal} with the peak at $k=k_{\rm d}$, the GW spectrum has a contribution from the $T_\Phi^{\rm eMD}(k<k_{\rm d})$ piece for GW momenta $0<k<2k_{\rm d}$ and from the $T_\Phi^{\rm eMD}(k>k_{\rm d})$ piece for GW momenta $2k_{\rm d}>k>2k_{\rm UV}$.

\subsection{The gravitational-wave spectral abundance}\label{sec:GW_spectral_abundance}
Let us now study the GW spectral abundance given by~\cite{Maggiore_2000}
\beq\label{eq:omegaGWdef}
\Omega_\mathrm{GW}(k,\eta) = \frac{1}{12}\frac{k^2}{\mathcal{H}^2}\mathcal{P}_h(k,\eta)\,.
\eeq
Note that $\eta$ should be chosen as the time when all relevant modes have entered the Hubble horizon. Here, we choose $\eta=\eta_{\rm evap}$ when modes of our interest satisfying $k\eta_{\rm evap}\gg 1$ can freely propagate. Then, we use \Eq{eq:phplrd} to calculate the tensor power spectrum approximately. To do so, we study below in detail the integral over momenta in the tensor power spectrum, which can be expressed as follows
\begin{align}\label{eq:prdapp}
    \overline{\cal P}_{h,\rm lRD}=\frac{c_s^4x_{\rm evap}^8}{2^7{\bar x}^2}\int_0^\infty {\rm d}v\int_{|1-v|}^{1+v}{\rm d}u &\left[\frac{4v^2-(1+v^2-u^2)^2}{4uv}\right]^2{\cal P}_\Phi(vk){\cal P}_\Phi(uk)\nonumber\\
    &\times v^2u^2\;\overline{{\cal I}^2}(\bar x,u,v,x_{\rm evap})\,,
    \end{align}
where $\mathcal{P}_{\Phi}(k)$ and its scale dependence is given by Eqs.~(\ref{eq:P_Phi}), (\ref{eq:scale1}) and (\ref{eq:scale2}). ${\cal P}_{\Phi}$ can be parameterized by a power-law form with a piecewise spectral index, i.e. ${\cal P}_{\Phi}\propto k^n$. Since $\overline{{\cal I}^2}$ introduces a $k^{-2}$ dependence, one can readily observe that the integrand in \Eq{eq:prdapp} peaks at large $k$ for $n>-2$. Therefore, as a qualitative estimate, we expect the tensor power spectrum to peak around $k_{\text{d}}$ and $k_{\text{UV}}$ in the case shown in Eq.~(\ref{eq:scale2}) where $k_{\rm c}>k_{\rm d}$, and around only $k_{\text{UV}}$ in the case shown in Eq.~(\ref{eq:scale1}) where $k_{\rm c}<k_{\rm d}$. 
We then perform below an analytic approximation of the gravitational-wave spectral abundance in each regime of interest.

\subsubsection{$k_\mathrm{c}>k_\mathrm{d}$}
This regime corresponds to a relatively large $\tau_{\text{NL}}$, as depicted in the right panel of \Fig{fig:P_PBH_vs_tau_NL}. The Poisson and clustering terms dominate in the regions of $k_c < k < k_{\rm UV}$ and $k > k_c$ in the scalar power spectrum ${\cal P}_\Phi$, respectively. Thus, dropping $k$-dependent factors we can divide the integral as follows
\begin{eqnarray}
\label{eq:prdapp2}
    \overline{\cal P}_{h,\rm lRD}(k)&\propto& 
    \int_{v_{\rm evap}}^{v_{\rm d}} {\rm d}v\int_{{\rm max}(|1-v|,v_{\rm evap})}^{{\rm min}(1+v,v_{\rm d})}{\rm d}u\,v^{4/3}u^{4/3}[...]+\int_{v_{\rm d}}^{v_{\rm c}}\int_{{\rm max}(|1-v|,v_{\rm d})}^{{\rm min}(1+v,v_{\rm c})}v^{-8/3}u^{-8/3}\cr\cr
    &&+\int_{v_{\rm c}}^{v_{\rm UV}}\int_{{\rm max}(|1-v|,v_{\rm c})}^{{\rm min}(1+v,v_{\rm UV})}v^{1/3}u^{1/3}+(\text{other terms})\,,   
\end{eqnarray}
where $\left[\dots\right]$ is the remaining $u$ and $v$ dependence in the prefactor of the integrand in \Eq{eq:prdapp}, which we have omitted here and below, along with the integral measure. We define $v_{\rm evap}$, $v_{\rm d}$, $v_{\rm c}$ and $v_{\rm UV}$ as $v_{\rm evap}\equiv k_{\rm evap}/k, v_{\rm d} \equiv k_{\rm d}/k, v_{\rm c} \equiv k_{\rm c}/k$ and $v_{\rm UV} \equiv k_{\rm UV}/k$. We will also omit ``other terms'' below as they are always subdominant. 

According to the discussion above, for the scales deep inside the Hubble horizon at $\eta_{\rm evap}$ we primarily focus on the resonant approximation \Eq{eq:irdapp}. For a given tensor mode at wavenumber $2k$, its main contribution comes from the scalar modes in the neighborhood of the resonance, namely those around $k$. This implies that the tensor modes in the wavenumber range $2k_{\rm c}<k<2k_{\rm UV}$ is primarily sourced by the $k_{\rm c}<k<k_{\rm UV}$ scalar modes where the Poisson term dominates. Similarly, the tensor spectrum in the wavenumber range $2k_{\rm d}<k<2k_{\rm c}$ is mainly governed by the $k_{\rm d}<k<k_{\rm c}$ region of the scalar spectrum where the clustering term dominates, etc.

Consequently, the first and second terms of \Eq{eq:prdapp2} dominates the tensor power spectrum in the ranges $2k_{\rm evap}<k<2k_{\rm d}$ and $2k_{\rm d}<k<2k_{\rm c}$, respectively. In these ranges, the scalar power spectrum is strongly dominated by the clustering term. This gives a distinctive feature from the Gaussian case in the GW spectral shape, specifically, a peak located at around $k_{\rm d}$ as mentioned before. Combining (\ref{eq:irdapp}) and (\ref{eq:prdapp}) then substituting them into (\ref{eq:omegaGWdef}), we arrive at an approximate formula for the GW spectrum in these $k$ ranges: 
        \begin{eqnarray}\label{eq:kdkc}
            \Omega_{\rm GW,res}(2k_{\rm d}< k< 2k_{\rm c})\simeq \frac{2187\pi c_s^{25/3}(1-c_s^2)^2}{16364}\left(\frac{9}{2}\right)^{1/3}\left(\frac{k_{\rm d}}{k}\right)^{7/3}\left(\frac{k_{\rm d}}{k_{\rm evap}}\right)^{17/3}\cr\cr
            \times \left(\nu^4 \bar{\tau}_{\rm NL}{\cal P}_{\cal R}(k)\right)^2\int^{s_1(k)}_{-s_1(k)}\frac{(1-s^2)^2}{(1-c_s^2s^2)^{14/3}}{\rm d}s\,,\nonumber\\
        \end{eqnarray}
        where the subscript ``res'' represents resonant approximation and the integral limit $s_1(k)$ is given by
        \begin{align}\label{eq:s1}
            s_1(k)=\left\{
            \begin{aligned}
            &c_s^{-1}-2\tfrac{k_{\rm d}}{k} \qquad &\tfrac{c_s^{-1}-1}{2}\leq\tfrac{k_{\rm d}}{k}\leq\tfrac{c_s^{-1}}{2}\quad\text{and}\quad\tfrac{k_{\rm c}}k\geq\tfrac{1+c_s^{-1}}{2}\\
            &1\qquad &\tfrac{k_{\rm d}}{k}\leq\tfrac{c_s^{-1}-1}{2}\quad\text{and}\quad \tfrac{k_{\rm c}}{k}\geq\tfrac{1+c_s^{-1}}{2}\\
            &2\tfrac{k_{\rm c}}{k}-c_s^{-1}\qquad & \tfrac{1+c_s^{-1}}{2}\geq\tfrac{k_{\rm c}}{k}\geq\tfrac{c_s^{-1}}{2}\\
            &0\qquad &\tfrac{c_s^{-1}}{2}\geq\tfrac{k_{\rm c}}{k}
            \end{aligned}
            \right.\,,
            \end{align}

   and     
        \begin{eqnarray}
            \Omega_{\rm GW,res}(2k_{\rm evap}< k< 2k_{\rm d})\simeq\frac{\pi c_s^{1/3}(1-c_s^2)^2}{640000\times 6^{\frac{1}{3}}}\left(\frac{k}{k_{\rm evap}}\right)^{17/3}\left(\nu^4 \bar{\tau}_{\rm NL}{\cal P}_{\cal R}(k)\right)^2\cr\cr
            \qquad\qquad\qquad\qquad\qquad\times \int^{s_2(k)}_{-s_2(k)}\frac{(1-s^2)^2}{(1-c_s^2s^2)^{2/3}}{\rm d}s\,,\nonumber\\
        \end{eqnarray}
        where
        \begin{align}\label{eq:s2}
            s_2(k)=\left\{
            \begin{aligned}
            &c_s^{-1}-2\tfrac{k_{\rm evap}}{k} \qquad &\tfrac{c_s^{-1}-1}{2}\leq\tfrac{k_{\rm evap}}{k}\leq\tfrac{c_s^{-1}}{2}\quad\text{and}\quad\tfrac{k_{\rm d}}k\geq\tfrac{1+c_s^{-1}}{2}\\
            &1\qquad &\tfrac{k_{\rm evap}}{k}\leq\tfrac{c_s^{-1}-1}{2}\quad\text{and}\quad \tfrac{k_{\rm d}}{k}\geq\tfrac{1+c_s^{-1}}{2}\\
            &2\tfrac{k_{\rm d}}{k}-c_s^{-1}\qquad & \tfrac{1+c_s^{-1}}{2}\geq\tfrac{k_{\rm d}}{k}\geq\tfrac{c_s^{-1}}{2}\\
            &0\qquad &\tfrac{c_s^{-1}}{2}\geq\tfrac{k_{\rm d}}{k}
            \end{aligned}
            \right.\,.
            \end{align}
        Note that when $k > 2c_s k_{\rm c}$, as indicated by \Eq{eq:s1}, the gravitational wave signal exhibits a cutoff. This is simply because within this range the approximate expression \Eq{eq:kdkc} is no longer applicable and the last term in \Eq{eq:prdapp2} becomes dominant. 

Similarly, the third term in \Eq{eq:prdapp2} dominates the tensor power spectrum in the range 
        \begin{eqnarray}\label{eq:kckUVres}
        \Omega_{\rm GW,c,\rm res}(2k_{\rm c}<k<2k_{\rm UV})
        \simeq\frac{3c_s^{7/3}(1-c_s^2)^2}{2^{14}\pi}\left(\frac{9}{2}\right)^{1/3}\left(\frac{k}{k_{\rm evap}}\right)^{11/3}\left(\frac{k_{\rm UV}}{k_{\rm evap}}\right)^2
        \left(\frac{k_{\rm d}}{k_{\rm UV}}\right)^8\cr\cr\times\int_{-s_0(k)}^{s_0(k)} \frac{(1-s^2)^2}{(1-c_s^2s^2)^{5/3}}{\rm d}s\,,\nonumber\\
        \end{eqnarray}
        where \cite{Inomata:2019ivs}
        \begin{align}\label{eq:s0}
            s_0(k)=\left\{
            \begin{aligned}
            &1\qquad &\tfrac{k_{\rm c}}{k}\leq\tfrac{c_s^{-1}-1}{2}\quad\text{and}\quad \tfrac{k_{\rm UV}}{k}\geq\tfrac{1+c_s^{-1}}{2}\\
            &2\tfrac{k_{\rm UV}}{k}-c_s^{-1}\qquad & \tfrac{1+c_s^{-1}}{2}\geq\tfrac{k_{\rm UV}}{k}\geq\tfrac{c_s^{-1}}{2}\\
            &0\qquad &\tfrac{c_s^{-1}}{2}\geq\tfrac{k_{\rm UV}}{k}
            \end{aligned}
            \right.\,.
            \end{align} 
        We can find that the gravitational wave signal peaks around $k \sim 2c_s k_{\rm UV}$, beyond which the spectrum is rapidly suppressed until the cutoff at $k = 2k_{\rm UV}$, in agreement with \cite{Domenech:2020ssp}.

Regrading the IR tail of the GW signal, the large $u,v$ contribution \eqref{eq:irdapp2} becomes more relevant. Focusing on the first term in \Eq{eq:prdapp2}, we approximate the IR tail of the GW spectrum as 
\begin{eqnarray}
    \label{eq:kevakdLV1}
    \Omega_{\rm GW,LV}(2k_{\rm evap}<k<2k_{\rm d})\simeq \frac{c_s^4}{110000}\left(\frac{3}{2}\right)^{2/3}\left(\frac{k}{k_{\rm evap}}\right)\left(\frac{k_{\rm d}}{k_{\rm evap}}\right)^{11/3}\left(\nu^4 \bar{\tau}_{\rm NL}{\cal P}_{\cal R}(k)\right)^2\cr\cr
    \times\left[1-\frac{11}{10}\left(\frac{k}{k_{\rm d}}\right)^2+\frac{2^{1/3}9}{10}\left(\frac{k}{k_{\rm d}}\right)^{11/3}-\frac{11}{16}\left(\frac{k}{k_{\rm d}}\right)^4\right]\,,\nonumber\\
\end{eqnarray}
where the subscript ``LV'' represents the large $u,v$ approximation. Note that this formula gives a $\Omega_{\rm GW} \propto k$ scaling in the IR region where $k\sim k_{\rm evap}$. 
\subsubsection{$k_\mathrm{c}<k_\mathrm{d}$}
This regime corresponds to a very small $\tau_{\text{NL}}$, as depicted in the left panel of \Fig{fig:P_PBH_vs_tau_NL}, where the clustering term in the scalar power spectrum corresponds to the region between $k_{\rm evap}$ and $k_{\rm c}$, exhibiting a relatively small amplitude. As mentioned above, we expect the GW spectral abundance increases with $k$ and peaks at $\sim k_{\rm UV}$. 
Similar to the discussion of $k_\mathrm{d}<k_\mathrm{c}$ regime, dropping $k$-dependent factors we can divide the integral as follows 
\begin{eqnarray}
\label{eq:prdapp3}
    \overline{\cal P}_{h,\rm lRD}(k)&\propto&
    \int_{v_{\rm evap}}^{v_{\rm c}} {\rm d}v\int_{{\rm max}(|1-v|,v_{\rm evap})}^{{\rm min}(1+v,v_{\rm c})}{\rm d}u\,v^{4/3}u^{4/3}[...]+\int_{v_{\rm c}}^{v_{\rm d}}\int_{{\rm max}(|1-v|,v_{\rm c})}^{{\rm min}(1+v,v_{\rm d})}v^{-8/3}u^{-8/3}\cr\cr
    &&+\int_{v_{\rm d}}^{v_{\rm UV}}\int_{{\rm max}(|1-v|,v_{\rm d})}^{{\rm min}(1+v,v_{\rm UV})}v^{1/3}u^{1/3}
    \,,    
\end{eqnarray}
while also neglecting terms like in \Eq{eq:prdapp2}. Taking into account the scaling of ${\cal P}_\Phi$, one can find the last term in \Eq{eq:prdapp3} remains dominant, as the integrand in this term features a positive power of momentum. This implies that the tensor modes in these regions are primarily induced by scalar modes near $k_{\rm UV}$, in contrast to the regime where $k_{\rm c}>k_{\rm d}$. Following the same approach we obtain the GW spectral abundance
\begin{eqnarray}\label{eq:kdkUVres}
        \Omega_{\rm GW,\rm res}(2k_{\rm d}<k<2k_{\rm UV})
        \simeq\frac{3c_s^{7/3}(1-c_s^2)^2}{2^{14}\pi}\left(\frac{9}{2}\right)^{1/3}\left(\frac{k}{k_{\rm evap}}\right)^{11/3}\left(\frac{k_{\rm UV}}{k_{\rm evap}}\right)^2
        \left(\frac{k_{\rm d}}{k_{\rm UV}}\right)^8\cr\cr
        \times\int_{-s_0(k)}^{s_0(k)} \frac{(1-s^2)^2}{(1-c_s^2s^2)^{5/3}}{\rm d}s\,,\nonumber\\
        \end{eqnarray}
    where $s_0(k)$ is the same as in \Eq{eq:s0}, as in both regimes, GWs at large $k$ are primarily sourced from the Poisson term. 
    The large $u,v$ contribution is 
    \begin{align}\label{eq:kevapkUVLV}
        \Omega_{\rm GW,\rm LV}(2k_{\rm evap}<k<2k_{\rm UV})=\frac{729c_s^4}{81920\pi^2}&\left(\frac{3}{2}\right)^{2/3}\left(\frac{k_{\rm evap}}{k_{\rm UV}}\right)^{10/3}\left(\frac{k_{\rm d}}{k_{\rm evap}}\right)^8\left(\frac{k}{k_{\rm UV}}\right)\,,
    \end{align}
    which dominates the IR tail of the GW spectral abundance, providing also a $\Omega_{\rm GW}\propto k$ scaling. 

\begin{figure}[H]
\begin{center}
\includegraphics[width = 0.85\textwidth]{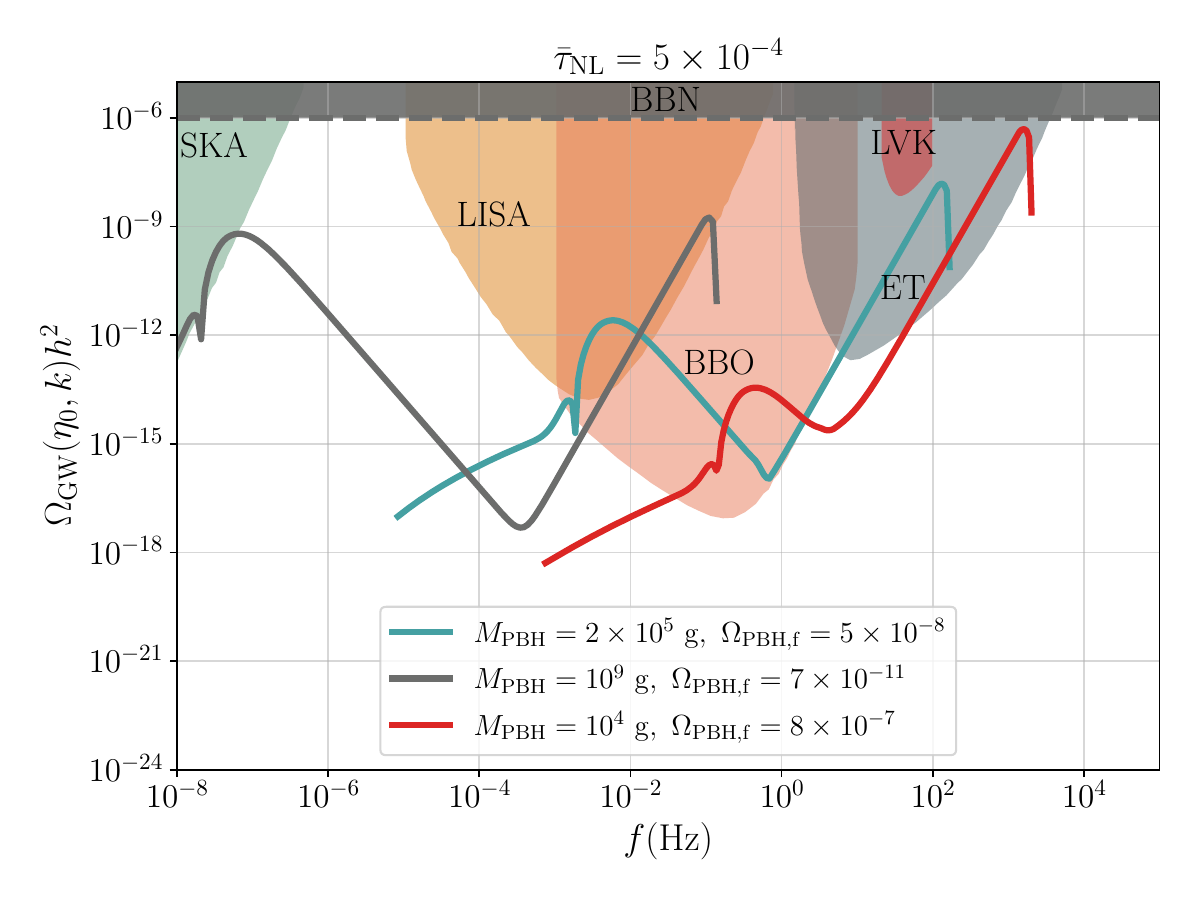}
\includegraphics[width = 0.85\textwidth]{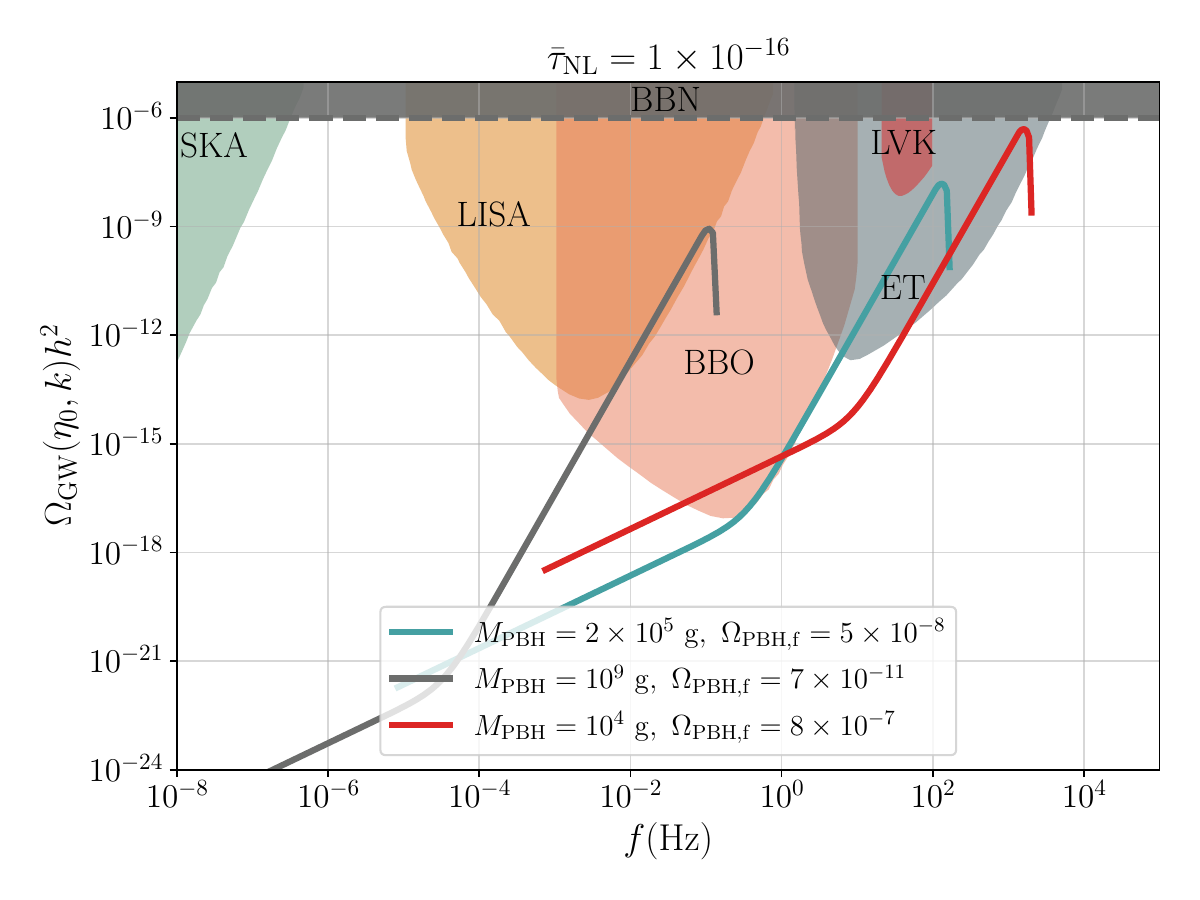}
\caption{{The GW signals for several parameter choices (as shown in the figure) for the cases where $k_{\rm c} > k_{\rm d}$ (Uper Panel) and $k_{\rm c} < k_{\rm d}$ (Lower Panel), corresponding to relatively large and extremely small $\tau_{\rm NL}$ values, respectively. The sensitivity curves of GW experiments, including LIGO-VIRGO-KAGRA (LVK) \cite{KAGRA:2021kbb}, LISA \cite{LISA:2017pwj, Karnesis:2022vdp}, ET \cite{Maggiore:2019uih}, SKA \cite{Janssen:2014dka}, and BBO \cite{Harry:2006fi}, are also overlaid for comparison.}}
\label{fig:OmegaGW}
\end{center}
\end{figure}

Considering the cosmic evolution that follows, we can determine the GW spectral abundance at present time, $\eta = \eta_0$, which reads
\beq
%$
\Omega_\mathrm{GW}(\eta_0,k)h^2 = 
%0.4\left(\frac{g_*(T_\mathrm{evap})}{106.75}\right)^{-1/3}
c_{\rm g} \Omega_{\mathrm{r},0} h^2 \Omega_\mathrm{GW}(\eta_\mathrm{evap},k)\,,
%$
\eeq
where $c_\mathrm{g} = 0.4 \left({g_*(T_\mathrm{evap})}/{106.75}\right)^{-1/3}$,
$\Omega_{\mathrm{r},0}h^2 \simeq 4\times 10^{-5}$~\cite{Planck:2018vyg},
and $\Omega_\mathrm{GW}(\eta_\mathrm{evap},k)$ is the GW spectral abundance evaluated at the time $\eta_\mathrm{evap}$, which we calculated above. With the same parameter choice as \Fig{fig:P_PBH_vs_tau_NL}, we show in \Fig{fig:OmegaGW} the GW signal induced by the scalar perturbations sourced from PBH isocurvature fluctuations. 

The upper panel in Figure \ref{fig:OmegaGW} corresponds to the case where $k_{\rm c} > k_{\rm d}$ and $\tau_{\rm NL}$ is not extremely small. One can observe a distinctive bi-peak structure in the GW spectral shape. The high-frequency peak corresponds to the Poisson distribution at small scales ($k \sim k_{\rm UV}$), while the low-frequency peak arises from the clustering effect on relatively larger scales due to primordial non-Gaussianities. Notably, even for a small $\tau_{\rm NL} = 5 \times 10^{-4}$, the low-frequency peak is clearly distinguishable. For the parameter choices shown in the figure, the bi-peak structures fall within the sensitivity ranges of future GW detectors, including LISA \cite{LISA:2017pwj, Karnesis:2022vdp}, ET \cite{Maggiore:2019uih}, SKA \cite{Janssen:2014dka}, and BBO \cite{Harry:2006fi}. 
The lower panel in Figure \ref{fig:OmegaGW} corresponds to the case where $k_{\rm c} < k_{\rm d}$ and $\tau_{\rm NL}$ is extremely small. As mentioned above, in this regime, the induced GWs are primarily sourced by the Poisson fluctuations at $k \sim k_{\rm UV}$. The result recovers the Gaussian case as discussed in Ref.~\cite{Domenech:2020ssp}.

\section{Conclusions}
\label{sec:conclusion}
PBHs constitute an active field of research since one can probe with these objects a wide variety of physical phenomena related to the physics of the early Universe up to the formation of large-scale structures~\cite{Carr:2020gox}. In particular, PBHs are associated with numerous GW signals~\cite{LISACosmologyWorkingGroup:2023njw}, being detectable by current and future GW experiments. In the present work, we focused on the GW signal induced at second order in cosmological perturbation theory by fluctuations in the number density of PBHs, or equivalently by PBH energy density fluctuations which are isocurvature in nature~\cite{Papanikolaou:2020qtd,Domenech:2020ssp,Papanikolaou:2022chm}. 

In particular, we derived firstly the two-point PBH correlation function in presence of local-type non-Gaussianities by generalising the work of~\cite{Suyama:2019cst} in the case of scale-dependent non-linearity parameters $f_\mathrm{NL}$, $g_\mathrm{NL}$ and $\tau_\mathrm{NL}$ finding at the end that the power spectrum of the initial isocurvature PBH energy density perturbations can be recast as 
\beq
{\cal P}_{\delta_\mathrm{PBH}}(k)={\cal P}_{\delta_\mathrm{PBH},\mathrm{NG}}(k)+{\cal P}_{\delta_\mathrm{PBH},\mathrm{Poisson}}(k), 
\eeq
where ${\cal P}_{\delta_\mathrm{PBH},\mathrm{Poisson}}(k)$ is the Gaussian Poissonian PBH matter power spectrum given by 
\beq
{\cal P}_{\delta_\mathrm{PBH},\mathrm{Poisson}}(k) = \frac{2}{3\pi}\left(\frac{k}{k_{\rm UV}}\right)^3,
\eeq
while ${\cal P}_{\delta_\mathrm{PBH},\mathrm{NG}}(k)$ is the non-Gaussian correction to ${\cal P}_{\delta_\mathrm{PBH}}(k)$ reading as
\beq
{\cal P}_{\delta_\mathrm{PBH},\mathrm{NG}}(k) =  \left(\frac{4\nu}{9\sigma_R}\right)^4{\cal P}_{\mathcal{R}}(k)\int \frac{\mathrm{d}^3p_1 \mathrm{d}^3p_2}{(2\pi)^6}\tau_\mathrm{NL}(p_1,p_2)W^2_\mathrm{local}(p_1)W^2_\mathrm{local}(p_2)P_{\mathcal{R}}(p_1)P_{\mathcal{R}}(p_2).
\eeq
Note that the non-Gaussian term ${\cal P}_{\delta_\mathrm{PBH},\mathrm{NG}}(k)$, being proportional to the primordial inflationary curvature power spectrum, gives rise to PBH clustering on large scales.

Having extracted the non-Gaussian PBH matter power spectrum we derived followingly both numerically and analytically the gravitational wave signal induced by the initially isocurvature PBH energy density perturbations. Notably, as it can be seen from our analytic expressions for the GW spectral abundance in \Sec{sec:GW_spectral_abundance}, one is met with two GW peaks in the presence of local-type scale-dependent primordial non-Gaussianities. Remarkably, the low-frequency peak is always present whenever there exists a non-zero local $\tau_\mathrm{NL}$ at the PBH scale with its amplitude being proportional to the square of $\tau_\mathrm{NL}$. interestingly enough, we found as well that this double-peaked GW signal can lie within the frequency bands of future GW detectors, namely that of LISA, ET, SKA and BBO, hence rendering our GW signal potentially detectable by GW experiments and promoting it as a novel portal probing the primordial non-Gaussianity.

Our analyses focused to sharply-peaked primordial power spectra around the PBH scale, leading to monochromatic PBH mass functions. One in principle can generalise our analysis to extended PBH mass functions and deriving at the end the effect on the induced GW signal [See here~\cite{Papanikolaou:2022chm} for the case of initial Gaussian PBH matter perturbations.]. In addition, we need to note that in order to be model-agnostic we studied the effect of a scale-dependent non-linearity parameter $\tau_\mathrm{NL}$ by phenomenologically parametrising it with a log-normal distribution peaked at the scale of PBH formation. Our formalism is however valid for whatever scale-dependence of $\tau_\mathrm{NL}$, which will in principle depend on the specifics of the underlying cosmological setup.

\begin{acknowledgments}
We thank Shi Pi, Teruaki Suyama, Shuichiro Yokoyama, and Yingli Zhang for valuable discussions. 
This work is supported in part by National Key R\&D Program of China (2021YFC2203100), by NSFC (12433002, 12261131497), by CAS young interdisciplinary innovation team (JCTD-2022-20), by 111 Project (B23042), by Fundamental Research Funds for Central Universities, by CSC Innovation Talent Funds, by USTC Fellowship for International Cooperation, by USTC Research Funds of the Double First-Class Initiative. It is also supported in part by the JSPS KAKENHI grants No. 20H05853 and 24K00624. TP and ENS acknowledge the contribution of the LISA CosWG and the COST Actions  CA18108 ``Quantum Gravity Phenomenology in the multi-messenger approach''  and CA21136 ``Addressing observational tensions in cosmology with systematics and fundamental physics (CosmoVerse)''. TP acknowledges also the support of INFN Sezione di Napoli \textit{iniziativa specifica} QGSKY as well as financial support from the Foundation for Education and European Culture in Greece.  
\end{acknowledgments}

\appendix
\section{PBH matter power spectrum}\label{app:PBH matter power spectrum}
Following the formalism developed in~\cite{Matarrese:1986et} and \cite{Suyama:2019cst} we obtain a general formula for the PBH two-point correlation function in presence of non-Gaussianities. 

We introduce firstly the local smoothed primordial fluctuations $\theta_{\rm local}({\bm x})$ defined as 
\begin{equation}
\theta_{\rm local} ({\bm x}) \equiv  \int d^3 y\, W_{\rm local} ({\bm x} - {\bm y}) ~\theta ({\bm y}),
\label{eq:local}
\end{equation}
where the window function $W_{\rm local}({\bm x})$ is a smoothing 
function with a smoothing scale $R$, which also removes scales larger than the scale $R$. By assuming that the criterion for PBH formation is given by
\begin{equation}
\theta_{\rm local} \geq \theta_{\rm th}\,,
\end{equation}
the probability to form a PBH at a point ${\bm x}$ is given by
\begin{equation}
P_1({\bm x}) =  \int [D\theta] P[\theta] \int_{\theta_{\rm th}}^{\infty} d\alpha\, \delta_D (\theta_{\rm local}({\bm x}) - \alpha),
\end{equation}
where $P[\theta]$ is the probability distribution function for the primordial fluctuations, $\theta ({\bm x})$. One can define as well the probability to have formed PBHs at points ${\bm x}_1$ and ${\bm x}_2$, which will read as
\begin{equation}
P_2 ({\bm x}_1,\,{\bm x}_2) = \int [D\theta] P[\theta] \int_{\theta_{\rm th}}^{\infty} d\alpha_1\, \delta_D (\theta_{\rm local}({\bm x}_1) - \alpha_1)
\int_{\theta_{\rm th}}^{\infty} d\alpha_2\, \delta_D (\theta_{\rm local}({\bm x}_2) - \alpha_2).
\end{equation}
By using the expression for the one-dimensional Dirac delta function given as
\begin{equation}
\delta_D (x) = \int \frac{d\phi}{2 \pi} e^{i \phi x},
\end{equation}
and the expression for the local smoothed fluctuations given by Eq.\,(\ref{eq:local}), we can expand the probability of PBH formation in terms of the $n$-point correlation functions of the primordial fluctuations by applying the path-integral approach developed by \cite{Matarrese:1986et}. Thus, the probability that a PBH forms at a point ${\bm x}$ can be recast as
\begin{equation}
    P_1({\bm x})=\frac{1}{\sqrt{2 \pi}} \int_\nu^{\infty} d w \exp \left[\sum_{n=3}^{\infty} \frac{(-1)^n}{n !} \frac{\xi_{\text {local }(c)}^{(n)}({\bm x})}{\sigma_{\text {local }}^n} \frac{d^n}{d w^n}\right] e^{-\frac{w^2}{2}}\,,
    \label{eq:p1}
    \end{equation}
while the probability to have two PBHs at points ${\bm x_1}$ and ${\bm x_2}$ will read as
\begin{equation}
    \begin{aligned}
    P_2\left(\boldsymbol{x}_1, \boldsymbol{x}_2\right)=\frac{1}{2 \pi} \int_\nu^{\infty} d w_1 & \int_\nu^{\infty} d w_2 \exp \left[\frac{\xi_{\text {local }(c)}^{(2)}\left(\boldsymbol{x}_1, \boldsymbol{x}_2\right)}{\sigma_{\text {local }}^2} \frac{\partial^2}{\partial w_1 \partial w_2}\right. \\
    & \left.+\sum_{n=3}^{\infty}(-1)^n \sum_{m=0}^n \frac{\xi_{\text {local }(c), m}^{(n)}\left(\boldsymbol{x}_1, \boldsymbol{x}_2\right) / \sigma_{\text {local }}^n}{m !(n-m) !} \frac{\partial^n}{\partial w_1^m \partial w_2^{n-m}}\right] e^{-\frac{w_1^2+w_2^2}{2}}.
    \end{aligned}
    \label{eq:p2}
    \end{equation}
where $\xi^{(n)}_{\text {local }(c)}$ denotes the $n$th moment of the local smoothed energy density field $\theta_{\rm local}$ being defined as 
\begin{equation}
    \xi_{{\rm local} (c)}^{(n)} (\boldsymbol{x}) \equiv \int d^3 y_1 d^3 y_2 \cdots d^3 y_n \, \xi_{\theta (c)} ({\bm y}_1,{\bm y}_2, \cdots, {\bm y}_n)
    \prod^n_{r=1} W_{\rm local} ({\bm x} - {\bm y}_r).
    \end{equation}
The $\xi_{\text {local }(c), m}^{(n)}$ moment on the other hand is defined as
\begin{eqnarray}
    \xi_{{\rm local}(c),m}^{(n)}\left(\boldsymbol{x}_1, \boldsymbol{x}_2\right)  & \equiv &\xi_{{\rm local}(c)}^{(n)}(
    \underbrace{{\bm x}_1, {\bm x}_1, \cdots, {\bm x}_1}_{{\rm total}~m},\underbrace{{\bm x}_2, {\bm x}_2, \cdots {\bm x}_2}_{{\rm total}~n-m}) \cr\cr
    & \equiv & \int d^3 y_1 d^3 y_2 \cdots d^3 y_n \, \xi_{\theta (c)} ({\bm y}_1,{\bm y}_2, \cdots, {\bm y}_n) \cr\cr
    && \qquad\qquad \times
    \prod^m_{r_1 = 1} W_{\rm local} ({\bm x}_1 - {\bm y_{r_1}})\prod^n_{r_2 = m+1} W_{\rm local}({\bm x}_2 - {\bm y}_{r_2})\,. \nonumber \\
    \end{eqnarray}
    
In the scenario we are considering, the distance between $\bm x_1$ and $\bm x_2$ is much larger than the smoothing scale, $|{\bm x}_1-{\bm x}_2|\gg R$, thus $\xi_{{\rm local} (c)}^{(n)} (\boldsymbol{x})$ and $\xi_{{\rm local}(c),m}^{(n)}\left(\boldsymbol{x}_1, \boldsymbol{x}_2\right)$ may differ significantly. We take 2-order moment for an example. The 2-order moment at one point $x$ can be expressed as 
\begin{eqnarray}\label{eq:2m1}
    \xi_{{\rm local} (c)}^{(2)} (\boldsymbol{x})=\int \frac{{\rm d}k}{k}W_{\rm local}(k)^2\mathcal{P}_\theta(k)\,,
\end{eqnarray}
where $P_\theta(k)$ is the power spectrum of the primordial fluctuation $\theta$, while the 2-order moment respect to two distant points $x_1$ and $x_2$ can be written as 
\begin{eqnarray}\label{eq:2m2}
    \xi_{{\rm local} (c),1}^{(2)} (\boldsymbol{x}_1,\boldsymbol{x}_2)=\int \frac{{\rm d}k}{k}j_0(kr)W_{\rm local}(k)^2\mathcal{P}_\theta(k)\,,
\end{eqnarray}
where $r\equiv|{\bm x}_1-{\bm x}_2|$ and zero-order spherical Bessel function $j_0(kr)\equiv {\sin (kr)}/{(kr)}$. According to the definition of $W_{\text{local}}(k)$, it selectively retains modes with $k \sim R^{-1}$ from the integrals in \Eq{eq:2m1} and \Eq{eq:2m2} while filtering out other modes. From \Eq{eq:2m2}, $ \xi_{{\rm local} (c),1}^{(2)} \sim j_0(r/R)\xi_{{\rm local} (c)}^{(2)} =j_0(r/R)\sigma_{\rm local}^2$, where $j_0(r/R)\ll1$ when $r\gg R$. Therefore, we denote $\xi_{{\rm local} (c),1}^{(2)} $ as $\xi_{{\rm local} (c),1}^{(2)} \equiv\epsilon \sigma_{\rm local}^2$, where $\epsilon \ll 1 $ when $\bm x_1$ and $\bm x_2$ are far apart. 
Similarly, for higher-order moments, a hierarchical scaling $\xi_{{\rm local} (c)}^{(n)}\propto \sigma_{\rm local}^{2n-2}$ allows us to approximately represent $\xi_{{\rm local} (c),m}^{(n)}$ as $\xi_{{\rm local} (c),m}^{(n)}\sim \epsilon\sigma_{\rm local}^{2n-2}$. 

It is important to stress here that in order to obtain the approximate expressions for $P_1$ and $P_2$ from \eqref{eq:p1} and \eqref{eq:p2}, we have considered (a) the high peak limit, and (b) the weak non-Gaussian limit. In the high peak limit, $\nu\equiv\theta_\mathrm{th}/\sigma_\mathrm{local}\gg 1$, an assumption which be well justified for nearly monochromatic PBH mass functions. In the weak non-Gaussian limit, we assume that $\nu^n\xi^{(n)}_{{\rm local}(c)}/\sigma_{\rm local}^n \ll 1$ for $n \geq 3 $, $\xi^{(n)}_{{\rm local}(c),m}/\sigma_{\rm local}^n \ll 1$ for $n \geq 3 $, which requires $\sigma_{\rm local}\ll \nu^{-3}$. 
Additionally, we assume that $\epsilon \ll \nu^{-2}$, as mentioned earlier, then the condition $\nu^n\xi^{(n)}_{{\rm local}(c),m}/\sigma_{\rm local}^n \ll 1$ for $n \geq 2 $ are well satisfied. With these assumptions, we can expand the exponential in \Eq{eq:p1} and \Eq{eq:p2} as a decreasing series and obtain an approximate result by truncating at a certain order. 
Therefore, for convenience, we impose that $\epsilon$ is of the same order as $\sigma_{\rm local}$ and then expand up to the order of  $\sigma_{\rm local}^3$: 
\begin{eqnarray}
    \label{eq:P1app}
    P_1 (\bm x) &\simeq&
    \frac{e^{-\nu^2/2}}{\sqrt{2 \pi} \nu} \Bigg\{ 1 + \frac{{S}^{(3)}}{3!}\nu^3 + \Bigg[\frac{{S}^{(4)}}{4!}\nu^4+\frac{1}{2!}\left(\frac{{S}^{(3)}}{3!}\right)^2\nu^6\Bigg]\cr\cr
    &&\qquad\qquad+\Bigg[\frac{{S}^{(5)}}{5!}\nu^5+\frac{{S}^{(3)}{S}^{(4)}}{3!4!}\nu^7+\frac{1}{3!}\left(\frac{{S}^{(3)}}{3!}\right)^3\nu^9\Bigg]+\dots
        \Bigg\}\,,
    \end{eqnarray}
For $P_2({\bm x}_1, {\bm x}_2)$, we can split it into two parts
\begin{eqnarray}
    \label{eq:P2app}
    P_2 ({\bm x}_1, {\bm x}_2) &\simeq&
    \frac{e^{-\nu^2}}{{2 \pi} \nu} \Bigg\{ 1 + \frac{{2S}^{(3)}}{3!}\nu^3 + 2\Bigg[\frac{{S}^{(4)}}{4!}\nu^4+\left(\frac{{S}^{(3)}}{3!}\right)^2\nu^6\Bigg]\cr\cr
    &&\qquad\qquad+2\Bigg[\frac{{S}^{(5)}}{5!}\nu^5+\frac{2{S}^{(3)}{S}^{(4)}}{3!4!}\nu^7+\frac{2}{3!}\left(\frac{{S}^{(3)}}{3!}\right)^3\nu^9\Bigg]+\dots
         \Bigg\}\cr\cr
    &&
    +  \frac{e^{-\nu^2}}{2 \pi \nu^2}\Bigg\{{{S}^{(2)}}_{,1}\nu^2+\Bigg[{{S}^{(3)}}_{,1}\nu^3+\frac{1}{2}\left({{S}^{(2)}}_{,1}\right)^2\nu^4+
    \frac{2{{S}^{(2)}}_{,1}{S}^{(3)}}{3!}\nu^5\Bigg]\cr\cr
    &&\qquad\qquad+\Bigg[\frac{2{{S}^{(4)}}_{,1}}{3!}\nu^4+\frac{{{S}^{(4)}}_{,2}}{2!2!}\nu^4+\frac{2{{S}^{(2)}}_{,1}{{S}^{(3)}}_{,1}}{2!}\nu^5+\frac{1}{3!}\left({{S}^{(2)}}_{,1}\right)^3\nu^6\Bigg]+\dots
     \Bigg\}\cr\cr
    &=&{P_1}^2+\frac{e^{-\nu^2}}{2 \pi \nu^2}\Bigg\{{{S}^{(2)}}_{,1}\nu^2+\Bigg[{{S}^{(3)}}_{,1}\nu^3+\frac{1}{2}\left({{S}^{(2)}}_{,1}\right)^2\nu^4+
    \frac{1}{3}{{{S}^{(2)}}_{,1}{S}^{(3)}}\nu^5\Bigg]\cr\cr
    &&\qquad\qquad+\Bigg[\frac{1}{3}{{S}^{(4)}}_{,1}\nu^4+\frac{1}{4}{{{S}^{(4)}}_{,2}}\nu^4+{{S}^{(2)}}_{,1}{{S}^{(3)}}_{,1}\nu^5+\frac{1}{6}\left({{S}^{(2)}}_{,1}\right)^3\nu^6\Bigg]+\dots
     \Bigg\}
     \,.\nonumber\\
    \end{eqnarray}
where $S^{(n)}\equiv \xi_{{\rm local}(c)}^{(n)}/\sigma_{\rm local}^n\propto \sigma_{\rm local}^{n-2}$ and ${S^{(n)}}_{,m}\equiv \xi_{{\rm local}(c),m}^{(n)}/\sigma_{\rm local}^n\propto \epsilon\sigma_{\rm local}^{n-2}\equiv \sigma_{\rm local}^{n-1}$.

At the end, by using the truncated expressions for $P_1$ and $P_2$, i.e. Eqs.~(\ref{eq:P1app}) and (\ref{eq:P2app}), the two-point correlation function of the PBHs defined as 
\begin{equation}
    \xi_{\rm PBH} ({\boldsymbol x}_1, {\boldsymbol x}_2)\equiv\frac{P_2({\bm x}_1,{\bm x}_2) }{P_1^2} - 1\,,
\end{equation}
can be evaluated up to the four-point correlation function of the local smoothed primordial overdensity field as
\begin{eqnarray}
    \xi_{\rm PBH} ({\boldsymbol x}_1, {\boldsymbol x}_2) &\sim& \frac{\nu^2}{\sigma_{\rm local}^2} \xi_{{\rm local}(c)}^{(2)}({\boldsymbol x}_1, {\boldsymbol x}_2) \cr\cr
    &&
    + \frac{1}{2} \frac{\nu^3}{\sigma_{\rm local}^3} \left( \xi^{(3)}_{{\rm local}(c)} ({\boldsymbol x}_1, {\boldsymbol x}_1, {\boldsymbol x}_2)  +  ({\boldsymbol x}_1 \leftrightarrow {\boldsymbol x}_2) \right) \cr\cr
    &&  + \frac{1}{4} \frac{\nu^4}{\sigma_{\rm local}^4} \xi^{(4)}_{{\rm local}(c)} ({\boldsymbol x}_1, {\boldsymbol x}_1, {\boldsymbol x}_2,{\boldsymbol x}_2) \cr\cr
    && 
    + \frac{1}{6} \frac{\nu^4}{\sigma_{\rm local}^4} \left( \xi^{(4)}_{{\rm local}(c)} ({\boldsymbol x}_1, {\boldsymbol x}_2, {\boldsymbol x}_2,{\boldsymbol x}_2) +  ({\boldsymbol x}_1 \leftrightarrow {\boldsymbol x}_2) \right)\cr\cr
    &&
    +(\text{nonlinear terms})\,, \nonumber\\
    \label{eq:PBHcorr}
    \end{eqnarray}
where the ``nonlinear terms'' in the last line are expressed as  
\begin{eqnarray}
    \label{eq:nonlinear}
    (\text{nonlinear terms}) &=& \frac{\nu^4}{2\sigma^4}\left(\xi_{\text {local }(c)}^{(2)}({\boldsymbol x}_1, {\boldsymbol x}_2)\right)^2+\frac{\nu^6}{6\sigma^6}\left(\xi_{\text {local }(c)}^{(2)}({\boldsymbol x}_1, {\boldsymbol x}_2)\right)^3\cr\cr
    &&
    +\frac{\nu^5}{6\sigma^5}\xi_{\text {local }(c)}^{(2)}({\boldsymbol x}_1, {\boldsymbol x}_2)\left(\xi^{(3)}_{{\rm local}(c)} ({\boldsymbol x}_1, {\boldsymbol x}_1, {\boldsymbol x}_1)+({\boldsymbol x}_1 \leftrightarrow {\boldsymbol x}_2)\right)\cr\cr
    &&
    +\frac{\nu^5}{2\sigma^5}\xi_{\text {local }(c)}^{(2)}({\boldsymbol x}_1, {\boldsymbol x}_2)\left(\xi^{(3)}_{{\rm local}(c)} ({\boldsymbol x}_1, {\boldsymbol x}_1, {\boldsymbol x}_2)+({\boldsymbol x}_1 \leftrightarrow {\boldsymbol x}_2)\right)\,. \nonumber\\
\end{eqnarray}

Hereafter, we choose a specific variable for $\theta_{\rm local}$. The smoothed density perturbation on comoving slices, $\delta_R({\bm x})$,  is a natural choice for the {\it local} primordial fluctucation $\theta_{\rm local}$ if we refer the fluctuation $\theta({\bm x})$ to the primordial curvature perturbation on comoving slices ${\cal R}$.

We introduce now the PBH power spectrum which is defined as
\begin{equation}\label{eq:PBH_power_spectrum_def}
    P_\text{PBH}(k)\equiv \int {\rm d}^3 r \,\xi_\text{PBH}({\bm r})e^{-i\bm{k\cdot r}}.
\end{equation}
Thus, substituting Eqs.(\ref{eq:PBHcorr}) and (\ref{eq:nonlinear}) into \Eq{eq:PBH_power_spectrum_def} we obtain that
\begin{eqnarray}
    P_{\delta_{\rm PBH}}(k) &\simeq& \frac{\nu^2}{\sigma^2_R}W_R(k)^2 P_{\delta} (k) \cr\cr
    && + \frac{\nu^3}{\sigma_R^3}\int \frac{{\rm d}^3 p}{(2 \pi)^3} W_R(p) W_R(|{\bm k} - {\bm p}|) W_R(k)
    B_\delta ({\bm p},-{\bm k},{\bm k} - {\bm p}) \cr\cr
    && + \frac{1}{3}\frac{\nu^4}{\sigma_R^4}\int \frac{{\rm d}^3 p_1 {\rm d}^3 p_2}{(2 \pi)^6}\cr\cr
    && \quad \times W_R(p_1)W_R(p_2) W_R(|{\bm k} - {\bm p}_1 - {\bm p}_2|) W_R(k)
    T_\delta ({\bm p}_1,{\bm p}_2, -{\bm k}, {\bm k} - {\bm p}_1 - {\bm p}_2) \cr\cr
    &&
    + \frac{1}{4}\frac{\nu^4}{\sigma_R^4}
    \int \frac{{\rm d}^3 p_1 {\rm d}^3 p_2}{(2 \pi)^6} \cr\cr
    && \quad  \times
    W_R(p_1)W_R(p_2) W_R(|{\bm k} - {\bm p}_1|)W_R(|{\bm p}_2 + {\bm k}|)
    T_\delta ({\bm p}_1,{\bm p}_2, {\bm k}-{\bm p}_1, -{\bm k} - {\bm p}_2)\cr\cr
    &&
    +P_{\delta_{\rm PBH}}^\text{NL}(k)~, \nonumber\\
    \label{eq:Ppbh1}
    \end{eqnarray}
where $P_\text{PBH}^\text{NL}(k)$ corresponds to the non-linear terms in \Eq{eq:nonlinear}, expressed as
\begin{eqnarray}
    P_{\delta_{\rm PBH}}^\text{NL}(k) &=& \frac{\nu^4}{2\sigma_R^4}\int \frac{{\rm d}^3 p}{(2\pi)^3}W_R(p)^2 W_R(|\bm{k-p}|)^2 P_\delta(p) P_\delta (|\bm{k-p}|) \cr\cr
    &&
    +\frac{\nu^6}{6\sigma_R^6}\int \frac{{\rm d}^3 p_1 {\rm d}^3 p_2}{(2 \pi)^6}\cr\cr
    && \quad \times W_R(p_1)^2 W_R(p_2)^2 W_R(|{\bm k} - {\bm p}_1 - {\bm p}_2|)^2 P_\delta(p_1)P_\delta(p_2)P_\delta(|\bm{k-p_1-p_2}|)\cr\cr
    && 
    +\frac{\nu^5}{3\sigma_R^5}\xi_{\text {local }(c)}^{(3)}  W_R(k)^2 P_\delta(k)\cr\cr
    &&
    +\frac{\nu^5}{\sigma_R^5}\int \frac{{\rm d}^3 p_1 {\rm d}^3 p_2}{(2 \pi)^6}\cr\cr
    && \quad \times W_R(p_1)W_R(p_2) W_R(|\bm{p_1}+\bm{p_2}|)W_R(|{\bm k} - {\bm p}_1 - {\bm p}_2|)^2 P_\delta(|{\bm k} - {\bm p}_1 - {\bm p}_2|)\cr\cr
    && \quad  \times B_\delta(\bm{p_1},\bm{p_2},-\bm{p_1}-\bm{p_2})\,. \nonumber\\
    \label{eq:Ppbhnl1}
\end{eqnarray}
Here, $W_R$ is the window function that smooths the modes with wavelength much smaller than the smoothing scale $R$. 

According to the discussion from Sec.\,\ref{sec:PBH_matter_PS_non_Gaussian}, $P_\delta,B_\delta,~$and $T_\delta$ are respectively
given in terms of the non-linearity parameters and the power spectrum of the primordial curvature perturbations as
\begin{eqnarray}
P_\delta (k) &=&  \left(\frac{4}{9}\right)^2 (kR)^4  P_{{\mathcal R}}(k), \cr\cr
B_\delta ({\bm k}_1, {\bm k}_2, {\bm k}_3) &=& \left(\frac{4}{9}\right)^3 (k_1R)^2 (k_2R)^2 (k_3R)^2\;\frac{6}{5}f_{\rm NL}({\bm k}_1, {\bm k}_2, {\bm k}_3)\cr\cr
&& \times\left[ P_{{\mathcal R}}(k_1) P_{{\mathcal R}}(k_2) + 2~{\rm perms.} \right], \cr\cr
T_\delta ({\bm k}_1, {\bm k}_2, {\bm k}_3, {\bm k}_4) &=& 
\left(\frac{4}{9}\right)^4 (k_1R)^2 (k_2R)^2 (k_3R)^2(k_4R)^2 \cr\cr
&& \times \bigg\{
\frac{54}{25}g_{\rm NL}({\bm k}_1, {\bm k}_2, {\bm k}_3, {\bm k}_4) [ P_{{\mathcal R}}(k_1) P_{{\mathcal R}}(k_2) P_{{\mathcal R}}(k_3)  + 3~{\rm perms.} ]  \cr\cr
&&\quad   + \tau_{\rm NL}({\bm k}_1, {\bm k}_2, {\bm k}_3, {\bm k}_4) [  P_{{\mathcal R}}(k_1) P_{{\mathcal R}}(k_2) P_{{\mathcal R}}(|{\bm k}_1 + {\bm k}_3|) + 11~{\rm perms.} ] \bigg\}\,.\nonumber\\
\label{eq:Deltaspec}
\end{eqnarray}
Substituting Eq.~(\ref{eq:Deltaspec}) into Eq.\,(\ref{eq:Ppbh1}), we have
\begin{eqnarray}
P_{\delta_{\rm PBH}}(k) &\simeq& \left(\frac{4\nu}{9\sigma_R}\right)^2 W_{\rm local}(k)^2 P_{\cal R} (k) \cr\cr
&&
+ \frac{6}{5} \left(\frac{4\nu}{9\sigma_R} \right)^3 
W_{\rm local}(k) \cr\cr
&& \times
\int \frac{{\rm d}^3 p}{(2 \pi)^3} f_{\rm NL}(k, p, |{\bm k} - {\bm p}|) W_{\rm local}(p) W_{\rm local}(|{\bm k} - {\bm p}|)\cr\cr
&& \quad \times 
\left[ 2 P_{\cal R}(p) P_{\cal R}(k) + P_{\cal R}(p) P_{\cal R}(|{\bm k} - {\bm p}|) \right] \cr\cr
&& + \frac{18}{25}\left(\frac{4\nu}{9\sigma_R} \right)^4 W_{\rm local}(k) \cr\cr
&& \times
\int \frac{{\rm d}^3 p_1 {\rm d}^3 p_2}{(2 \pi)^6}g_{\rm NL}(k, p_1, p_2, |{\bm k} - {\bm p}_1 - {\bm p}_2|)\cr\cr
&& \quad\quad \times W_{\rm local}(p_1)W_{\rm local}(p_2) W_{\rm local}(|{\bm k} - {\bm p}_1 - {\bm p}_2|)\cr\cr
&& \quad\quad \times \left[ 3 P_{\cal R}(k)P_{\cal R}(p_1)P_{\cal R}(p_2) + P_{\cal R}(p_1)P_{\cal R}(p_2)P_{\cal R}(|{\bm k} - {\bm p}_1 - {\bm p}_2|) \right] \cr\cr
&&
 + \frac{1}{3}\left(\frac{4\nu}{9\sigma_R} \right)^4 W_{\rm local}(k) \cr\cr
&& \times
\int \frac{{\rm d}^3 p_1 {\rm d}^3 p_2}{(2 \pi)^6}\tau_{\rm NL}(k, p_1, p_2, |{\bm k} - {\bm p}_1 - {\bm p}_2|)\cr\cr
&& \quad\times W_{\rm local}(p_1)W_{\rm local}(p_2) W_{\rm local}(|{\bm k} - {\bm p}_1 - {\bm p}_2|) \cr\cr
&& \quad\times \left[ 6 P_{\cal R}(k)P_{\cal R}(p_1)P_{\cal R}(|{\bm p}_1 + {\bm p}_2|) + 6 P_{\cal R}(p_1)P_{\cal R}(p_2)P_{\cal R}(|{\bm k} - {\bm p}_1 |) \right] \cr\cr
&&
+ \frac{54}{25}\left(\frac{4\nu}{9\sigma_R}\right)^4
 \cr\cr
&& \times \int \frac{{\rm d}^3 p_1 {\rm d}^3 p_2}{(2 \pi)^6}g_{\rm NL}(p_1, p_2, |{\bm k} + {\bm p}_1|,|{\bm p}_2 - {\bm k}|) \cr\cr 
&&\qquad \times W_{\rm local}(p_1)W_{\rm local}(p_2) W_{\rm local}(|{\bm k} + {\bm p}_1|)W_{\rm local}(|{\bm p}_2 - {\bm k}|)\cr\cr
&& \qquad \times  P_{\cal R}(p_1) P_{\cal R}(p_2) P_{\cal R}(|{\bm k}+{\bm p}_1|) \cr\cr
&&
+ \frac{1}{4}\left(\frac{4\nu}{9\sigma_R}\right)^4 \cr\cr
&& \times 
\int \frac{{\rm d}^3 p_1 {\rm d}^3 p_2}{(2 \pi)^6}\tau_{\rm NL}(p_1, p_2, |{\bm k} + {\bm p}_1|, |{\bm p}_2 - {\bm k}|)\cr\cr
&& \times W_{\rm local}(p_1)W_{\rm local}(p_2) W_{\rm local}(|{\bm k} + {\bm p}_1|)W_{\rm local}(|{\bm p}_2 - {\bm k}|) 
 \cr\cr
&& \times \left[ 4 P_{\cal R}(p_1)P_{\cal R}(p_2) P_{\cal R}(k) \right.\cr\cr
&& \quad \left.
+ 4 P_{\cal R}(p_1) \left( P_{\cal R}(p_2) P_{\cal R}(|{\bm k} + {\bm p}_1 - {\bm p}_2|) + P_{\cal R} (|{\bm k} + {\bm p}_1|)P_{\cal R}(|{\bm p}_1 + {\bm p}_2|) \right) \right]\cr\cr
&& +P_{\delta_{\rm PBH}}^\text{NL}(k)\,, \nonumber\\
\label{eq:Ppbh2}
\end{eqnarray}
where
\begin{eqnarray}
    P_{\delta_{\rm PBH}}^\text{NL}(k) &=& \frac{1}{2}\left(\frac{4\nu}{9\sigma_R} \right)^4 \int \frac{{\rm d}^3 p}{(2\pi)^3} W_{\rm local}(p)^2  W_{\rm local}(|\bm{k-p}|)^2 P_{\cal R}(p) P_{\cal R} (|\bm{k-p}|) \cr\cr
    &&
    +\frac{1}{6}\left(\frac{4\nu}{9\sigma_R} \right)^6\int \frac{{\rm d}^3 p_1 {\rm d}^3 p_2}{(2 \pi)^6} W_{\rm local}(p_1)^2W_{\rm local}(p_2)^2 W_{\rm local}(|{\bm k} - {\bm p}_1 - {\bm p}_2|)^2\cr\cr
    && \qquad \times P_{\cal R}(p_1)P_{\cal R}(p_2)P_{\cal R}(|\bm{k-p_1-p_2}|)\cr\cr
    &&
    +\frac{6}{5} \left(\frac{4\nu}{9\sigma_R} \right)^5\int \frac{{\rm d}^3 p_1 {\rm d}^3 p_2}{(2 \pi)^6}f_{\rm NL}(p_1, p_2, |\bm{p_1}+\bm{p_2}|)\cr\cr
    && \qquad\times W_{\rm local}(p_1)W_{\rm local}(p_2) W_{\rm local}(|\bm{p_1}+\bm{p_2}|)W_{\rm local}(|{\bm k} - {\bm p}_1 - {\bm p}_2|)^2 \cr\cr
    &&\qquad\times P_{\cal R}(|{\bm k} - {\bm p}_1 - {\bm p}_2|)  [ P_{\cal R}(p_1) P_{\cal R}(p_2)\cr\cr
    && \qquad\qquad\times + P_{\cal R}(p_2) P_{\cal R}(|\bm{p_1}+\bm{p_2}|)+ P_{\cal R}(|\bm{p_1}+\bm{p_2}|) P_{\cal R}(p_1) ]\cr\cr
    &&
    +\frac{\nu^3}{3\sigma_R^3}\left(\frac{4\nu}{9\sigma_R} \right)^2\xi_{\text {local }(c)}^{(3)}  W_{\rm local}(k)^2 P_{\cal R}(k)
    \,, \nonumber\\
    \label{eq:PpbhNL2}
    \end{eqnarray}
where $W_{\rm local}(k) \equiv (kR)^2 \,W_R(k)$, and $\xi^{(3)}_{\text{local}(c)}$ is a constant independent with scales. For super-Hubble correlation of PBHs, i.e. for $kR\ll 1$ limit, noting that we can take $W_{\rm local}(k)\rightarrow 0$ and $|\bm{k+p}|\rightarrow p$, \Eq{eq:Ppbh2} becomes 
\begin{eqnarray}\label{eq:Appbhng}
    { P}_{\delta_\mathrm{PBH}}(k)\simeq&&\left(\frac{4\nu}{9\sigma_R}\right)^4{ P}_{\mathcal{R}}(k)\int \frac{\mathrm{d}^3p_1 \mathrm{d}^3p_2}{(2\pi)^6}\tau_\mathrm{NL}W^2_\mathrm{local}(p_1)W^2_\mathrm{local}(p_2)P_{\mathcal{R}}(p_1)P_{\mathcal{R}}(p_2)\cr\cr
    &&+(k\textrm{-independent terms})
    \,,\nonumber\\
\end{eqnarray}
where the $k$-independent terms are expressed as 
\begin{eqnarray}
\label{eq:kinde}
&&(k\textrm{-independent terms})\cr\cr
 &&\quad= \frac{1}{2}\left(\frac{4\nu}{9\sigma_R} \right)^4 \int \frac{{\rm d}^3 p}{(2\pi)^3} W_{\rm local}(p)^4 P_{\cal R}(p)^2 \cr\cr
    && \qquad
    +\frac{1}{6}\left(\frac{4\nu}{9\sigma_R} \right)^6\int \frac{{\rm d}^3 p_1 {\rm d}^3 p_2}{(2 \pi)^6}\cr\cr
    && \qquad\quad \times W_{\rm local}(p_1)^2W_{\rm local}(p_2)^2 W_{\rm local}(|{\bm p}_1 + {\bm p}_2|)^2 P_{\cal R}(p_1)P_{\cal R}(p_2)P_{\cal R}(|\bm{p_1+p_2}|)\cr\cr
    && \qquad
    +\frac{6}{5} \left(\frac{4\nu}{9\sigma_R} \right)^5\int \frac{{\rm d}^3 p_1 {\rm d}^3 p_2}{(2 \pi)^6}\cr\cr
    &&\qquad\quad\times f_{\rm NL}(p_1,p_2,|{\bm p_1+\bm p_2}|)\cr\cr
    && \qquad\quad\times W_{\rm local}(p_1)W_{\rm local}(p_2) W_{\rm local}(|\bm{p_1}+\bm{p_2}|)^3 \cr\cr
    &&\qquad\quad\times P_{\cal R}(|{\bm p}_1 + {\bm p}_2|)  \left[ P_{\cal R}(p_1) P_{\cal R}(p_2) + P_{\cal R}(p_2) P_{\cal R}(|\bm{p_1}+\bm{p_2}|)+ P_{\cal R}(|\bm{p_1}+\bm{p_2}|) P_{\cal R}(p_1)  \right]\cr\cr
    &&\qquad
+ \frac{54}{25}\left(\frac{4\nu}{9\sigma_R}\right)^4\int \frac{{\rm d}^3 p_1 {\rm d}^3 p_2}{(2 \pi)^6}g_{\rm NL}(p_1, p_2,p_1,p_2)W_{\rm local}(p_1)^2W_{\rm local}(p_2)^2 P_{\cal R}(p_1)^2 P_{\cal R}(p_2) \cr\cr
&&\qquad
+\left(\frac{4\nu}{9\sigma_R}\right)^4
\int \frac{{\rm d}^3 p_1 {\rm d}^3 p_2}{(2 \pi)^6}\tau_{\rm NL}(p_1,p_2,p_1,p_2) \cr\cr
&& \quad\qquad\times W_{\rm local}(p_1)^2W_{\rm local}(p_2)^2 
 \cr\cr
&& \quad \quad \qquad\times \left[
P_{\cal R}(p_1) \left( P_{\cal R}(p_2) P_{\cal R}(|{\bm p}_1 - {\bm p}_2|) + P_{\cal R} (p_ 1)P_{\cal R}(|{\bm p}_1 + {\bm p}_2|) \right) \right]
    \,. \nonumber\\
\end{eqnarray}
As one can observe, terms involving the non-linearity parameters $f_{\rm NL}, g_{\rm NL}$ and $\tau_{\rm NL}$ appear in the $k$-independent terms, which can be considered as the modification in PBH abundance, as mentioned at the beginning of Sec.\,\ref{sec:PBH_matter_PS_non_Gaussian}.

We need to mention that, given the exponential sensitivity of the PBH abundance to the tail of the distribution for the underlying density field, even slight deviations from Gaussianity can significantly modify the PBH density. This, in turn, affects the amplitude of the PBH matter power spectrum (${\cal P}_{\delta_{\rm PBH}} \propto \Omega_{\rm PBH}$). This modification is illustrated by the non-Gaussian correction terms in \Eq{eq:kinde}. 
However, in our work, we choose to set the initial abundance of PBHs in the presence of non-Gaussianities, $\Omega_{\rm PBH,f}$, as a given parameter instead of finding the complete and explicit relation between the correlation function of high peaks in the density field and non-linearity parameters, which is beyond the scope of our work. Consequently, we employ the same form as in \Eq{eq:ppoi} to calculate the Poisson term, ${\cal P}_{\delta_\mathrm{PBH},\mathrm{Poisson}}$, in \Eq{eq:ppbhng}, where $k_{\rm UV}$ is related to the given $\Omega_{\rm PBH,f}$. 
It is noteworthy that the clustering effect could result in a discrepancy between the breakdown scale of the PBH fluid description, $r_{\rm UV}$, and the mean separation distance, $\bar{r}$. Nonetheless, at least for purposes of characterizing the amplitude of the Poisson term and the related GW signal, it is safe to use $k_{\rm UV}=a/\bar{r}$ in our calculations.

\section{The late radiation-dominated era transfer function}
We shall define the transfer function from the isocurvature fluctuation to the Newtonian potential $\Phi$ by 
\begin{equation}\label{eq:tfdef}
    \Phi_{\rm lRD}(k,\eta)=T_\Phi(k,\eta)S(k). 
\end{equation}

In order to obtain the transfer function $T_{\Phi}(k,\eta)$ during the lRD era after the PBH evaporation, one needs to solve \Eq{eq:Phi_evolution} under the instantaneous evaporation approximation. At the end,  one can obtain an exact solution of $\Phi$ during the lRD era, 
~\cite{Domenech:2020ssp}
\begin{equation}\label{eq:philrd}
    \Phi_{\rm lRD}^{\rm instant}(k,\eta)=\frac{1}{c_s k\overline{\eta}}\left[C_1 j_1(c_s k\overline{\eta})+C_2 y_1(c_s k\overline{\eta})\right]~; \quad \overline{\eta}\equiv\eta-\eta_{\rm evap}/2~,
\end{equation}
where $c_s=1/\sqrt{3}$ is the radiation sound speed and the re-scaled conformal time $\eta\to\overline{\eta}$ is defined as $\eta\to\overline{\eta}\equiv\eta-\eta_{\rm evap}/2$ whereas the constants $C_1$ and $C_2$ are given by
\begin{equation}\label{eq:c1c2}
    C_1=-\frac{1}{8}\Phi_{\rm eMD}(c_sk\eta_{\rm evap})^3 y_2(c_s k\eta_{\rm evap}/2)\,,\quad C_2=\frac{1}{8}\Phi_{\rm eMD}(c_sk\eta_{\rm evap})^3 j_2(c_s k\eta_{\rm evap}/2)\,, 
\end{equation}
and were derived by requiring continuity of $\Phi$ and its first derivative at $\eta=\eta_{\rm evap}$. $j_i$ and $y_i$ in \Eq{eq:philrd} and \Eq{eq:c1c2} are the first and second kind of $i$-order spherical Bessel functions, respectively. At this point, we need to mention that $\Phi_\mathrm{eMD}=T_{\Phi,\rm eMD}(k)S$ is the constant value of $\Phi$ during the eMD era driven by PBHs~\cite{Kohri:2018awv}. Finally, plugging $C_1$ and $C_2$ into \Eq{eq:philrd} and also taking into account the suppression factor given by \Eq{eq:supfactor}, $T_{\Phi}$ reads as 
\begin{equation}\label{eq:TlRD}
    T_{\Phi}(k,\eta)= S_\Phi^2
 \frac{c_s^2(k\eta_{\rm eva})^3}{8k\overline{\eta}}\big[j_2(c_s k\eta_{\rm evap}/2)y_1(c_sk\overline{\eta})-y_2(c_s k\eta_{\rm evap}/2)j_1(c_sk\overline{\eta})\big]\,T_{\Phi,\rm eMD}(k)\,,
\end{equation}
%where $\mathcal{S}_\Phi(k)$ is the suppression to $\Phi$ given by Eq.\,(\ref{eq:supfactor}). 
where $T_{\Phi,\rm eMD}$ is given by Eq.~\eqref{eq:Phi_eMD}. 
For the scales of our interest, i.e. scales deep inside the Hubble horizon at evaporation $x_{\rm evap}=k\eta_{\rm evap}\gg 1$, \Eq{eq:TlRD} can be approximated as 
\begin{equation}\label{eq:T_Phi_lRD}
    T_{\Phi,{\rm lRD}}(k, \eta)= S_{\Phi}^2(k)\left(\frac{x_{\rm evap}}{2\bar{x}}\right)^2\cos\left[c_s(x-x_{\rm evap})\right]~,
\end{equation}
where $x\equiv k\eta$. Note that for the modes deep inside the Hubble horizon, $\mathcal{H}^{-1}T_{\Phi}'(\eta_{\rm evap})\sim (k\eta_{\rm evap})T_{\Phi}(\eta_{\rm eva})\gg T_{\Phi}(\eta_{\rm evap})$, equally $\mathcal{H}^{-1}\Phi_{\rm lRD}'\gg \Phi_{\rm lRD}$ at the time of evaporation. One then can clearly see that the large derivative term $\mathcal{H}^{-1}\Phi_{\rm lRD}'$ is the one that gives the dominant contribution in the source term \Eq{eq:Source:def} for $k>2k_{\rm evap}$. 

\section{Analytical approximation for calculation of induced GWs}
Let us turn now into the integral over momenta in the tensor power spectrum, which can be expressed as follows
\begin{align}\label{eq:appph}
    \overline{\cal P}_{h,\rm lRD}(k)&=8\int_0^\infty {\rm d}v\int_{|1-v|}^{1+v}{\rm d}u \left[\frac{4v^2-(1+v^2-u^2)^2}{4uv}\right]^2{\cal P}_S(vk){\cal P}_S(uk)\overline{I^2_{\rm lRD}}(u,v,x_{\rm evap})\nonumber\\
    &=\frac{c_s^4x_{\rm evap}^8}{2^7{\bar x}^2}\int_0^\infty {\rm d}v\int_{|1-v|}^{1+v}{\rm d}u \left[\frac{4v^2-(1+v^2-u^2)^2}{4uv}\right]^2{\cal P}_\Phi(vk){\cal P}_\Phi(uk)\nonumber\\
    &\qquad\qquad\qquad\qquad\qquad\qquad\times v^2u^2\overline{{\cal I}^2}(\bar x,u,v,x_{\rm evap})\,,
    \end{align}
where $\mathcal{P}_{\Phi}(k)$ and its scale dependence is give by \Eq{eq:scale1}. $\overline{{\cal I}^2}(\bar x,u,v,x_{\rm evap})$ in the two kinds of contribution mentioned in the main text can be expressed respectively as 
\begin{eqnarray}
    \label{eq:intres}
    &&\overline{{\cal I}^2_{\rm lRD,res}}(\bar x,u,v,x_{\rm evap})=\frac{1}{32}{\rm Ci}^2(|1-(u+v)c_s|x_{\rm evap}/2)\,,\\
    &&\overline{{\cal I}^2_{\rm lRD,LV}}(\bar x,u,v,x_{\rm evap})=\frac{1}{8}\Big[{\rm Ci}^2(x_{\rm evap}/2)+\big({\rm Si}(x_{\rm evap}/2)-\pi/2\big)^2\Big]\,.\label{eq:intLV}
\end{eqnarray}
Then, we can perform the calculation for the tensor power spectrum by combining different terms in Eq.~(\ref{eq:prdapp2}) and Eq.~(\ref{eq:prdapp}).

We first derive the resonant contribution from $k_{\rm d}<k<k_{\rm c}$ region of the scalar power spectrum corresponding to the third term in Eq.\,(\ref{eq:prdapp2}), which dominates the tensor spectrum in range $2k_{\rm d}<k<2k_{\rm c}$:
\begin{eqnarray}\label{eq:dphlrdkdkc}
    &&\overline{\cal P}_{h,\rm lRD}(k,\eta,x\gg 1)
    \simeq\frac{c_s^4}{2^{12}\, \bar{x}^2}\left(\frac{3}{2}\right)^{2/3}\left(\frac{k_{\rm evap}}{k}\right)^{4/3}\left(\frac{9}{8}\right)^4\left(\frac{k_{\rm d}}{k}\right)^{8} x_{\rm evap}^8 \left(\nu^4 \bar{\tau}_{\rm NL}{\cal P}_{\cal R}(k)\right)^2\cr\cr
    &&\times \int_{v_{\rm d}}^{v_{\rm c}} dv\int_{{\rm max}(|1-v|,v_{\rm d})}^{{\rm min}(1+v,v_{\rm c})}du\,\left[\frac{4v^2-(1+v^2-u^2)^2}{4u v}\right]^2(u v)^{-8/3}\,{\rm Ci}^2\left(|1-(u+v)c_s|x_{\rm evap}/2\right)
    \,.\nonumber\\
\end{eqnarray}
For the double integral in the above equation, we follow Ref.\,\cite{Domenech:2020ssp} to perform the integral by changing variables to 
\begin{eqnarray}
    &&y\equiv \left((u+v)c_s-1\right){x_{\rm evap}}/{2}\,,\\
    &&s\equiv u-v\,.
\end{eqnarray}
The Jacobian of the transformation is 
\begin{equation}
    |J|=\frac 1{c_sx_{\rm evap}}\,.
\end{equation}
Then the transformed integrand can be given by 
\begin{eqnarray}\label{eq:tfig}
    \frac{2^{16/3}}{c_s x_{\rm evap}} \frac{(1-z^2)^2(1-s^2)^2}{(z^2-s^2)^{14/3}} {\rm Ci}^2(|y|)\,.
\end{eqnarray}
Then, as a good approximation we evaluate the integrand at $y=0$ except for the divergent cosine integral\,\cite{Inomata:2019ivs}, 
where we used that 
\begin{equation}
    \int_{-\infty}^{\infty}{\rm Ci}^2(|y|)\,{\rm d}y=\pi\,.
\end{equation}
At the end, we obtain
\begin{eqnarray}
    \overline{\cal P}_{h,\rm lRD}(k,\eta,x\gg 1)&\simeq&\frac{c_s^4}{2^{12}\, \bar{x}^2}\left(\frac{3}{2}\right)^{2/3}\left(\frac{k_{\rm evap}}{k}\right)^{4/3}\left(\frac{9}{8}\right)^4\left(\frac{k_{\rm d}}{k}\right)^{8} x_{\rm evap}^8 \left(\nu^4 \bar{\tau}_{\rm NL}{\cal P}_{\cal R}(k)\right)^2\cr\cr
   &&\qquad\qquad\qquad\times \frac{{2^{16/3}\pi c_s^{13/3}(1-c_s^2)^2}}{x_{\rm evap}}\int_{-s_1(k)}^{s_1(k)}\frac{(1-s^2)^2}{(1-c_s^2s^2)^{14/3}}\,{\rm d}s\cr\cr
   &=&\frac{{\pi c_s^{25/3}(1-c_s^2)^2}}{2^{20/3}\, \bar{x}^2}\left(\frac{3}{2}\right)^{2/3}\left(\frac{k_{\rm evap}}{k}\right)^{4/3}\left(\frac{9}{8}\right)^4\left(\frac{k_{\rm d}}{k}\right)^{8} x_{\rm evap}^7 \cr\cr
   &&\qquad\qquad\qquad\times \left(\nu^4 \bar{\tau}_{\rm NL}{\cal P}_{\cal R}(k)\right)^2\int_{-s_1(k)}^{s_1(k)}\frac{(1-s^2)^2}{(1-c_s^2s^2)^{14/3}}\,{\rm d}s
   \,.\nonumber\\
\end{eqnarray}
Finally, we arrive at an approximate formula for the GW spectrum in the wavenumber range $2k_{\rm d}<k<2k_{\rm c}$ given by 
\begin{eqnarray}
    \Omega_{\rm GW,res}(2k_{\rm d}\ll k\ll 2k_{\rm c})\simeq \frac{1}{12}\times\frac{{\pi c_s^{25/3}(1-c_s^2)^2}}{2^{20/3}\, \bar{x}^2}\left(\frac{3}{2}\right)^{2/3}\left(\frac{k_{\rm evap}}{k}\right)^{4/3}\left(\frac{9}{8}\right)^4\left(\frac{k_{\rm d}}{k}\right)^{8} \left(\frac{2k}{k_{\rm evap}}\right)^7 \cr\cr
    \qquad\qquad\qquad\times \left(\nu^4 \bar{\tau}_{\rm NL}{\cal P}_{\cal R}(k)\right)^2\int_{-s_1(k)}^{s_1(k)}\frac{(1-s^2)^2}{(1-c_s^2s^2)^{14/3}}\,{\rm d}s\cr\cr
    \simeq\frac{2187\pi c_s^{25/3}(1-c_s^2)^2}{16364}\left(\frac{9}{2}\right)^{1/3}\left(\frac{k_{\rm d}}{k}\right)^{7/3}\left(\frac{k_{\rm d}}{k_{\rm evap}}\right)^{17/3}\qquad\qquad\qquad\cr\cr
    \qquad\qquad\qquad\times \left(\nu^4 \bar{\tau}_{\rm NL}{\cal P}_{\cal R}(k)\right)^2\int^{s_1(k)}_{-s_1(k)}\frac{(1-s^2)^2}{(1-c_s^2s^2)^{14/3}}{\rm d}s\,.\nonumber\\
\end{eqnarray}
In the wavenumber range $2k_{\rm evap}<k<2k_{\rm d}$, both resonant and large $u,v$ contributions are important. We consider first the resonant contribution. Combining the second term in Eq.~(\ref{eq:prdapp2}), Eq.~(\ref{eq:appph}) and Eq.~(\ref{eq:intres}), we can obtain
\begin{eqnarray}\label{eq:appkevakd}
    &&\overline{\cal P}_{h,\rm lRD}(k,\eta,x\gg 1)\simeq\frac{c_s^4}{2^{12}\, \bar{x}^2}\left(\frac{3}{2}\right)^{2/3}\left(\frac{k_{\rm evap}}{k}\right)^{4/3}\left(\frac{1}{5}\right)^4 x_{\rm evap}^8 \left(\nu^4 \bar{\tau}_{\rm NL}{\cal P}_{\cal R}(k)\right)^2\cr\cr
    &&\quad\quad\times \int_{v_{\rm d}}^{v_{\rm c}} dv\int_{{\rm max}(|1-v|,v_{\rm d})}^{{\rm min}(1+v,v_{\rm c})}du\,\left[\frac{4v^2-(1+v^2-u^2)^2}{4u v}\right]^2(u v)^{4/3}\,{\rm Ci}^2\left(|1-(u+v)c_s|x_{\rm evap}/2\right)\,.\nonumber\\
\end{eqnarray}
Then follow the same transformation we have the transformed integrand (with Jacobian) in above equation expressed as 
\begin{eqnarray}
    \frac{c_s^{-11/3}(1-c_s^2)^2}{2^{8/3} x_{\rm evap}}\frac{(1-s^2)^2}{(1-c_s^2s^2)^{2/3}}{\rm Ci}^2(|y|)\,.
\end{eqnarray}
The integral in Eq.\,(\ref{eq:appkevakd})
\begin{eqnarray}
    \frac{c_s^{-11/3}(1-c_s^2)^2}{2^{8/3} x_{\rm evap}}\int_{-s_2(k)}^{s_2(k)}\frac{(1-s^2)^2}{(1-c_s^2s^2)^{2/3}}{\rm d}s\,.
\end{eqnarray}
Substituting this into Eq.\,(\ref{eq:appkevakd}) we obtain
\begin{eqnarray}
    &&\overline{\cal P}_{h,\rm lRD}(k,\eta,x\gg 1)\simeq \frac{c_s^4}{2^{12}\, \bar{x}^2}\left(\frac{3}{2}\right)^{2/3}\left(\frac{k_{\rm evap}}{k}\right)^{4/3}\left(\frac{1}{5}\right)^4 x_{\rm evap}^8 \left(\nu^4 \bar{\tau}_{\rm NL}{\cal P}_{\cal R}(k)\right)^2\cr\cr
    &&\qquad\qquad\qquad\qquad\qquad\times \frac{\pi c_s^{-11/3}(1-c_s^2)^2}{2^{8/3} x_{\rm evap}}\int_{-s_2(k)}^{s_2(k)}\frac{(1-s^2)^2}{(1-c_s^2s^2)^{2/3}}{\rm d}s\cr\cr
    &&\qquad\qquad\qquad\qquad = \frac{\pi c_s^{1/3}(1-c_s^2)^2}{2^{7}\, \bar{x}^2}\left(\frac{3}{4}\right)^{2/3}\left(\frac{1}{5}\right)^4\left(\frac{k}{k_{\rm evap}}\right)^{17/3} \left(\nu^4 \bar{\tau}_{\rm NL}{\cal P}_{\cal R}(k)\right)^2\cr\cr
    &&\qquad\qquad\qquad\qquad\qquad \times\int_{-s_2(k)}^{s_2(k)}\frac{(1-s^2)^2}{(1-c_s^2s^2)^{2/3}}{\rm d}s\,.\nonumber\\
\end{eqnarray}
Finally, we arrive at an approximate formula for the GW spectrum in the wavenumber range $2k_{\rm evap}<k<2k_{\rm d}$ given by 
\begin{eqnarray}
    \Omega_{\rm GW,res}(2k_{\rm evap}\ll k\ll 2k_{\rm d})\simeq \frac{1}{12}\times\frac{\pi c_s^{1/3}(1-c_s^2)^2}{2^{7}}\left(\frac{3}{4}\right)^{2/3}\left(\frac{1}{5}\right)^4\left(\frac{k}{k_{\rm evap}}\right)^{17/3} \cr\cr
    \qquad\qquad\qquad\times\left(\nu^4 \bar{\tau}_{\rm NL}{\cal P}_{\cal R}(k)\right)^2\int_{-s_2(k)}^{s_2(k)}\frac{(1-s^2)^2}{(1-c_s^2s^2)^{2/3}}{\rm d}s\cr\cr
    =\frac{\pi c_s^{1/3}(1-c_s^2)^2}{640000\times 6^{1/3}}\left(\frac{k}{k_{\rm evap}}\right)^{17/3}\left(\nu^4 \bar{\tau}_{\rm NL}{\cal P}_{\cal R}(k)\right)^2\qquad\cr\cr
    \times \int_{-s_2(k)}^{s_2(k)}\frac{(1-s^2)^2}{(1-c_s^2s^2)^{2/3}}{\rm d}s
    \,.\nonumber\\
\end{eqnarray}

For the large $u,v$ part in $2k_{\rm evap}<k<2k_{\rm d}$, combining the second term in Eq.\,(\ref{eq:prdapp2}), Eq.\,(\ref{eq:appph}) and Eq.\,(\ref{eq:intLV}), we obtain
\begin{eqnarray}\label{eq:appkevakdLV}
    \overline{\cal P}_{h,\rm lRD}(k,\eta,x\gg 1)&\simeq&\frac{c_s^4}{2^{10}\, \bar{x}^2}\left(\frac{3}{2}\right)^{2/3}\left(\frac{k_{\rm evap}}{k}\right)^{4/3}\left(\frac{1}{5}\right)^4 x_{\rm evap}^8 \left(\nu^4 \bar{\tau}_{\rm NL}{\cal P}_{\cal R}(k)\right)^2\cr\cr
    &&\times \int_{v_{\rm d}}^{v_{\rm c}} dv\int_{{\rm max}(|1-v|,v_{\rm d})}^{{\rm min}(1+v,v_{\rm c})}du\,\left[\frac{4v^2-(1+v^2-u^2)^2}{4u v}\right]^2(u v)^{4/3}\cr\cr
    &&\qquad\qquad\qquad\qquad\times\Big[{\rm Ci}^2(x_{\rm evap}/2)+\big({\rm Si}(x_{\rm evap}/2)-\pi/2\big)^2\Big]\,.\nonumber\\
\end{eqnarray}
The integral of the momenta in the large $u,v$ contribution can be done as follows. We change variables to 
\begin{eqnarray}
    &&t\equiv u+v-1\,,\\
    &&s\equiv u-v\,,
\end{eqnarray}
where the Jacobian is 
\begin{equation}
    |J|=\frac{1}{2}\,.
\end{equation}
We follow \cite{Domenech:2020ssp} to set $s=0$, then the feasible region for $t+1$ is the interval swept on the $u$-axis when the line $ u = -v + t + 1 $ sweeps the intersection of the lines $ u = v $ and the integral domain given by the integral in Eq.\,(\ref{eq:appkevakdLV}). 
\begin{equation}
    2^{-11/3}\int_{0}^{\tfrac{2k_{\rm d}}{k}-1} \frac{(1-(1+t)^2)^2}{(1+t)^{4/3}}{\rm d}t \,.
\end{equation}
Substituting into Eq.\,(\ref{eq:appkevakdLV}) we obtain
\begin{eqnarray}
    \overline{\cal P}_{h,\rm lRD}(k,\eta,x\gg 1)\simeq\frac{c_s^4}{2^{41/3}\, \bar{x}^2}\left(\frac{3}{2}\right)^{2/3}\left(\frac{k_{\rm evap}}{k}\right)^{4/3}\left(\frac{1}{5}\right)^4 x_{\rm evap}^8 \left(\nu^4 \bar{\tau}_{\rm NL}{\cal P}_{\cal R}(k)\right)^2\cr\cr
    \quad\quad\times \int_{0}^{\tfrac{2k_{\rm d}}{k}-1} \frac{(1-(1+t)^2)^2}{(1+t)^{4/3}}{\rm d}t\times\Big[{\rm Ci}^2(x_{\rm evap}/2)+\big({\rm Si}(x_{\rm evap}/2)-\pi/2\big)^2\Big]\cr\cr
    =\frac{c_s^4}{2^{41/3}\, \bar{x}^2}\left(\frac{3}{2}\right)^{2/3}\left(\frac{k_{\rm evap}}{k}\right)^{4/3}\left(\frac{1}{5}\right)^4\left(\frac{2k}{k_{\rm evap}}\right)^8 \left(\nu^4 \bar{\tau}_{\rm NL}{\cal P}_{\cal R}(k)\right)^2\left(\frac{k_{\rm evap}}{k}\right)^2\cr\cr
    \times\int_{0}^{\tfrac{2k_{\rm d}}{k}-1} \frac{(1-(1+t)^2)^2}{(1+t)^{4/3}}{\rm d}t
    \,.\nonumber\\
\end{eqnarray}
the additional $(k_{\rm evap}/k)^2$ factor is from the expansion of the cosine and sine integrals for the large arguement limit. The integral in the above equation can be calculated:
\begin{equation}
\begin{aligned}
    &\int_{0}^{\tfrac{2k_{\rm d}}{k}-1} \frac{(1-(1+t)^2)^2}{(1+t)^{4/3}}{\rm d}t  \\
    &= \frac{3}{110}\left[2^{\frac{2}{3}}\left(80\left(\frac{k_{\rm d}}{k}\right)^{\frac{11}{3}}-88\left(\frac{k_{\rm d}}{k}\right)^{\frac{5}{3}} 
     - 55\left(\frac{k}{k_{\rm d}}\right)^{\frac{11}{3}}\right)+144\right] ~.
\end{aligned}
\end{equation}
Taking its leading-order term and substituting into the above equation yields the result in Eq.\,(\ref{eq:kevakdLV1}). However, due to the relatively small difference between $k_{\rm evap}$ and $k_{\rm d}$ in our chosen values (within two orders of magnitude), although the leading-order term dominates its scale, other terms still contribute to the absolute value. 
%\bibliography{ref.bib}

\bibliography{main_new}
\bibliographystyle{JHEP}

\end{document}